\documentclass[12pt,a4paper,american,refpage]{paper}
\usepackage[T1]{fontenc}
\usepackage[latin9]{inputenc}
\usepackage{color}
\usepackage{babel}
\usepackage{array}
\usepackage{amstext}
\usepackage{amssymb}
\usepackage{graphicx}
\usepackage{nomencl}
\providecommand{\printnomenclature}{\printglossary}
\providecommand{\makenomenclature}{\makeglossary}
\makenomenclature
\usepackage[unicode=true,pdfusetitle,
 bookmarks=true,bookmarksnumbered=false,bookmarksopen=false,
 breaklinks=false,pdfborder={0 0 0},pdfborderstyle={},backref=false,colorlinks=true]
 {hyperref}

\makeatletter

\pdfpageheight\paperheight
\pdfpagewidth\paperwidth

\newcommand{\noun}[1]{\textsc{#1}}
\providecommand{\tabularnewline}{\\}

\usepackage{a4wide}
\usepackage{epsfig}
\usepackage{color}
\usepackage{changebar}
\newcommand{\decibel}{\ensuremath{\mathrm{dB}}}

\usepackage{xspace}
\newcommand{\etal}{\emph{et al.}\xspace}
\newcommand{\ie}{\emph{i.e.}\xspace}

\newcommand{\viterbi}{Viterbi\xspace}

\AtBeginDocument{}

\makeatother

\begin{document}

\global\long\def\comma{\,,}%

\global\long\def\point{\,.}%

\global\long\def\test{\lessgtr}%
 
\title{Error-rate in Viterbi decoding of a duobinary signal in presence of
noise and distortions: theory and simulation}
\author{Henri Mertens and Marc Van Droogenbroeck}
\institution{Montefiore Institute, University of Liège, Belgium}

\maketitle
\textbf{Abstract.} The Viterbi algorithm, presented in 1967, allows
a maximum likelihood decoding of partial response codes. This study
focuses on the duobinary code which is the first member of this family
and has been specified for the digital part of television systems
recommended by International Organizations. Up to now the error-rate,
which is the main criterion of the performance, has been evaluated
by simulation. Although there exist theoretical bounds, these bounds
are not satisfactory for a channel such as broadcasting (by terrestrial
transmitters, cable networks or satellite) which is strongly impaired
by noise, and linear and non-linear distortions. Analytical methods,
verified by simulation, are presented here in order to evaluate the
theoretical and exact values of the error-rate, in the form of series
of numerical integrations, for a transmission in baseband or in radio-frequency
with quadriphase modulation (or AM/VSB for cable networks) and coherent
demodulation, in presence of noise and several distortions. This methodology
can be later extended to other partial response codes, to convolutional
codes and their concatenations. 
\begin{keywords}
Viterbi decoding, bit-error rate, duobinary code, error-rate, convolutional
code, partial response signaling, digital modulation, interference,
noise, synchronization error, non-linear distortion, echo
\end{keywords}
\newpage{}

\tableofcontents{}

\subsection*{\newpage}

\subsection*{Notations}

\subsubsection*{Main notations}

The following main notations are used in this paper: 
\begin{center}
\begin{tabular}{cl}
$a_{i}$ & input binary bits -1 or +1\tabularnewline
$b_{i}$ & precoded bits -1 or +1\tabularnewline
$D$ & delay operator of one bit\tabularnewline
$d_{i}$ & duobinary transmitted symbols -1, 0 or +1\tabularnewline
$r_{i}$ & duobinary symbols recognized (correctly or not) by the decoder\tabularnewline
$n_{i}$ & noise samples\tabularnewline
$S_{1}$, $S_{2}$ & survivors corresponding to the states -1 and +1\tabularnewline
$L$ & length of a duobinary sequence\tabularnewline
$f_{b}$, $T$ & bit frequency and bit period\tabularnewline
$x_{i}$ or $y_{i}$ & input level at the decoder\tabularnewline
$x(t)$ & input signal at the decoder\tabularnewline
$y(t)$ & interference at input of decoder\tabularnewline
$\varphi$ & phase error\tabularnewline
$\tau$ & synchronization error or echo delay\tabularnewline
$\beta$ & echo relative amplitude\tabularnewline
$\psi$ & echo phase\tabularnewline
\end{tabular}
\par\end{center}

\subsubsection*{Some special notations}
\begin{center}
\begin{tabular}{cl}
$\test$ & symbol for the following test: ``$<$'' OR ``$>$''\tabularnewline
$a_{1}(t),\,a_{2}(t)$ & data signals modulating the two carriers in quadrature\tabularnewline
$r_{1}(t),\,r_{2}(t)$ & data signals after demodulation, at the decoder input\tabularnewline
$x(t)=x_{r}(t)+j\,x_{i}(t)$ & modulated signal \emph{before} distortion\tabularnewline
$z(t)=z_{r}(t)+j\,z_{i}(t)$ & modulated signal \emph{after} distortion\tabularnewline
$\rho(t)$ and $\rho_{d}(t)$ & envelopes of $x(t)$ and $z(t)$\tabularnewline
$\theta(t)$ and $\theta(t)+\varphi(t)$ & phases of $x(t)$ and $z(t)$\tabularnewline
$A_{c}$ & amplitude of the carrier\tabularnewline
$\omega_{c}$ & carrier angular frequency\tabularnewline
$A_{s}$ & saturation level of the TWT\tabularnewline
$B_{\textrm{off}}$ & backoff (see the conventional definition)\tabularnewline
$C$ & carrier power in absence of non-linear distortion\tabularnewline
$N_{0}$ & one-sided power spectral density of the noise in RF\tabularnewline
$\sigma^{2}$ & noise variance at the decoder\tabularnewline
\end{tabular}
\par\end{center}

\newpage{}

\section{Introduction}

While already proposed in 1967, the rules of the \viterbi algorithm~\cite{Viterbi1967Error}
are still used in a number of digital communication and broadcasting
systems. Applications in these fields, as well as other applications
such as magnetic recording, and even the more recent ``turbo-codes''~\cite{Berrou1994Turbo}
include rules of the \viterbi algorithm for several forms of RF\nomenclature{RF}{Radio Frequency}
multiplexing, such as Orthogonal Frequency Division Multiplexing (OFDM\nomenclature{OFDM}{Orthogonal Frequency Division Multiplexing})
and several forms of modulation (QPSK\nomenclature{QPSK}{Quadrature Phase Shift Keying},
QPRS\nomenclature{QPRS}{Quadrature Partial-Response Signaling}, 16
or 64-QAM\nomenclature{QAM}{Quadrature Amplitude Modulation},~...).

In communication systems, the \viterbi algorithm is commonly used
for decoding partial response codes and convolutional codes and provides
a large improvement over classical threshold decoding. The final error-rate
for these systems are often obtained only by simulations.

The purpose of this paper is to present and to corroborate, by comparison
with simulation, a methodology to derive the theoretical and exact
value of the error-rate in the form of an analytical expression, in
the case of a channel with strong impairments due to additive Gaussian
noise but also due to important forms of linear and non-linear distortions.
Such a channel is typically a broadcasting channel with data transmission
by terrestrial transmitters, by cable networks or by satellite. For
example severe causes of impairments are echoes with long delay and
multipath propagation, echoes with short delay, and non-linear distortions
in the power stage amplifier, respectively in terrestrial broadcasting,
in cable networks, and in satellite transmission. 

The paper is deliberately restricted to the study of the duobinary
code which is the first member of the family of partial response codes
and mainly because it has been already specified for the digital part
of television systems recommended for satellite broadcasting by the
EBU\nomenclature{EBU}{European Broadcasting Union}, CCIR\nomenclature{CCIR}{Comité consultatif international pour la radio (now ITU-R)}
and ITU\nomenclature{ITU}{International Telecommunication Union}
(Mac/packet systems)~\cite{CCIR-Rec-650-1,CCIR-Rep-1073,CCIR-Rep-632,EBU1986-tech-doc-3268-E,ITU-BO-787,ITU-BO-1074-1,CCIR1988Publication}.
However the presented methodology can be later extended to other partial
response codes, to convolutional codes, and even to their concatenations.

This paper focuses on the evaluation of the final bit error-rate (BER\nomenclature{BER}{Bit Error Rate})
which, in data transmission, is the  main criterion for the system
performance. In the type of channel considered, it is generally assumed
that the final quality should still be satisfactory with a BER of
the order of $10^{-3}$ or even more, account being taken that, for
example in teletext, the most sensitive bytes are again protected
by an appropriate error-correcting code. This level of BER is unacceptable
in applications such as magnetic recording where the BER should be
of the order of $10^{-6}$ or even less. 

Theoretical upper bounds of the BER with \viterbi decoding have already
been given in the literature. In the case of decoding convolutional
codes such bounds are given for example in~\cite{Viterbi1967Error}
and~\cite[pp. 462-470]{Spilker1977Digital}. For the case of decoding
partial-response codes, similar bounds are described in~\cite{Forney1972Maximum,Kobayashi1971Correlative,UER-Doc-GTR5-101},
and by a different approach in~\cite{Altekar1999Error}. The upper
bound given in these two last references for the duobinary code will
be examined in more details in Section~\ref{subsec:Upper-bound-of}.
The same kind of bound for another system was developed by Tjhung
\etal~\cite{Tjhung1989Error}; the system studied was narrow band
FM\nomenclature{FM}{Frquency Modulation} where the binary input is
first coded in duobinary, then FM modulated, detected by discriminator
and finally duobinary decoded by the \viterbi algorithm (note that
for this system, the FM noise, including the ``click'' noise is
no longer Gaussian and the\noun{ }\viterbi algorithm is no more optimum).

But, while all these bounds give a good approximation of the actual
BER at high values of the signal to noise ratio (\ie for very low
BER), they depart from the exact values when the BER is low, or in
other words in presence of strong noise or interference which is our
main concern. 

The paper is organized as follows. Section~\ref{sec:Brief-review-of-viterbi}
briefly describes the key elements and terminology of the \viterbi
algorithm. Then, we present the duobinary coding and propose a theoretical
framework capable to provide performance figures of a baseband transmission
system that combines a duobinary coding and the \viterbi decoding
algorithm, respectively in Section~\ref{sec:Duobinary-code} and
Section~\ref{sec:Duobinary-theoretical-framework}. Finally we analyze
the impact of several types of interferences or distortions with quadriphase
modulation and coherent demodulation in Section~\ref{sec:Interferences},
namely a phase error in coherent demodulation, a synchronization error,
an echo in radio-frequency (with CQPRS\nomenclature{CQPRS}{Continuous Quadrature Partial Response Signaling modulation}
or AM/SSB\nomenclature{SSB}{Single-Sideband modulation} modulation
taken as an approximation of AM/VSB\nomenclature{VSB}{Vestigial Sideband modulation}),
and the non-linear distortion in a model of a traveling-wave tube
frequently used as the satellite power amplifier.

\section{Brief review of \viterbi's decoding algorithm\label{sec:Brief-review-of-viterbi}}

The algorithm proposed by \viterbi in 1967~\cite{Viterbi1967Error}
is a rather general algorithm permitting to solve, with a limited
number of calculations, a number of problems for searching the shortest
path in a graph called a \emph{trellis}. 

\begin{figure}[htbp]
\begin{centering}
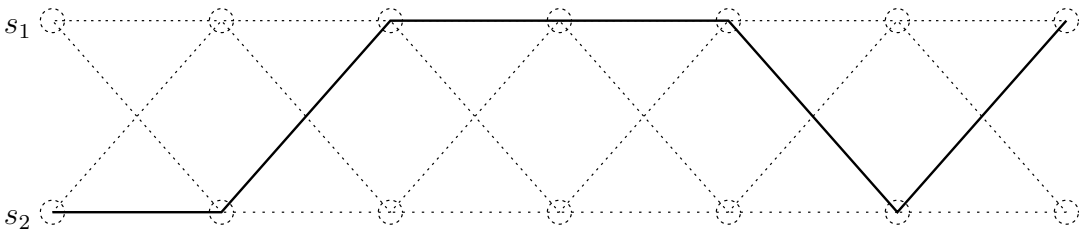
\par\end{centering}
\caption{Example of a path followed, from left to right, in a two nodes trellis.
The dotted lines correspond to other possible state transitions.\label{figure1}}
\end{figure}

Consider a linear system which can take $m$ different \emph{states}
at successive instants $1,\,2,\,\ldots,\,p-1,\,p$. These states are
the \emph{nodes} of the trellis. The evolution of the system is represented
by a trellis with $m$ nodes. An example of the path followed in a
trellis for the simple case $m=2$, which corresponds to the duobinary
code, is shown in Figure~\ref{figure1}; the two possible states
are noted $s_{1}$ and $s_{2}$. 

The transition from a node of order $p-1$ to a node of order $p$
is a \emph{branch} of the trellis. In order to evaluate the length
of the path followed in the trellis, it is necessary to define a \emph{metric},
\ie a rule giving the length of any branch.

Let us consider the $m$ possible states at instant $p-1$. To each
of these states, there is one and only one path having the shortest
metric from the origin. These $m$ shortest paths are called the \emph{survivors}
of order $p-1$. After this step, there are $m^{2}$ possible state-transitions
from order $p-1$ to order $p$, since from each node of order $p-1$
there are $m$ branches going to the $m$ nodes of order $p$.

When the algorithm is applicable, the two essential following results
have been demonstrated:
\begin{enumerate}
\item The shortest path from the origin to a node of order $p$ completely
contains the survivor of order $p-1$ of this path, the metrics being
additive, which is the case when the noise samples added to the signal
are statistically independent. Note that this assumption is true for
additive white Gaussian noise (AWGN\nomenclature{AWGN}{Additive White Gaussian Noise}).
\item All the survivors are converging to each other in the sense that the
probability of non-convergence has the limit zero after an infinite
number of steps; in practice however the convergence is obtained after
a finite number of state-transitions called the \emph{decoding constraint
length $L_{C}$} which for most systems is typically 20 to 30 bits.
In addition, in absence of impairment, the survivors converge to the
system states.
\end{enumerate}
Therefore, the calculation of the shortest path from the origin up
to any node is made by recurrence. At order $p-1$, the decoder keeps
in memory the $m$ survivors arriving at the nodes $p-1$. It then
computes the lengths of the $m^{2}$ transitions from order $p-1$
to order $p$ and adds the length of the $m$ transitions starting
from a node of order $p-1$ to the corresponding survivor of order
$p-1$. The result is the lengths of the $m$ possible paths from
the origin to the nodes of order $p$ but passing by a given node
of order $p-1$. Among these $m$ paths only the shortest one is retained.
This calculation is repeated for all nodes of order $p-1$, thus forming
the survivors of order $p$.

When applied to communication systems, the \viterbi algorithm no
longer gives \emph{hard decisions} (taken bit by bit) but \emph{soft}
\emph{decisions} based on the whole followed path. In presence of
white Gaussian noise it is then optimum in the sense of maximum a
posteriori likelihood. 

In the next section, we develop the duobinary signaling (with a brief
presentation of other partial response codes). Then we present analytical
methods for evaluating the theoretical BER in presence of noise and
distortions with duobinary and \viterbi decoding. 

\section{Duobinary code and partial response codes\label{sec:Duobinary-code}}

\subsection{General presentation}

A \emph{partial response code} is a weighted addition of $n$ successive
input data bits such as the coded symbol $d_{i}$ is given by 
\begin{equation}
d_{i}=k(h_{0}a_{i}+h_{1}a_{i-1}+h_{2}a_{i-2}+\ldots+h_{n}a_{i-n})\comma
\end{equation}
where $k$ is a scale factor and the coefficients $h$ define the
code. If we introduce the \emph{delay operator of one bit}, denoted
$D$, the code may equivalently be represented by the polynomial $P(D)$
defined as 

\begin{equation}
d_{i}=kP(D)=k(h_{0}+h_{1}D+h_{2}D^{2}+\ldots+h_{n}D^{n})\point
\end{equation}

\subsection{Duobinary code}

Duobinary coding (also referred to as Class~1) was proposed as a
mean to introduce some controlled amount of \emph{Inter-symbol} \emph{Interference}
(ISI\nomenclature{ISI}{Inter-Symbol Interference}) from the adjacent
bit rather than trying to eliminate it completely. If $a_{i}$ are
the binary input bits (0 or 1) and $d_{i}$ the duobinary symbols
(with the symmetrical values $-1,\,0,$ or $+1$) at time $t_{i}$,
the duobinary symbols are defined by the transverse filter~\cite{Kretzmer1966Generalization,Sklar1988Digital}
\begin{equation}
d_{i}=\frac{a_{i}+a_{i-1}}{2}=\frac{1}{2}\,(1+D)\point\label{equation1}
\end{equation}

If the binary signal has the form of short pulses $\delta(t)$, the
duobinary signal can also be obtained by convolution with the impulse
response of the \emph{duobinary filter} which is
\begin{equation}
G(f)=\cos\left(\frac{\pi f}{f_{b}}\right)\comma\label{equation2}
\end{equation}
$f_{b}$ being the bit frequency. The spectrum of the duobinary signal
is then cascaded with an ideal filter and strictly limited to the
Nyquist band $f_{b}/2$.

The duobinary filter can either be \emph{matched} (\ie split in equal
parts between the transmitter and the receiver) or \emph{unmatched}
(\ie entirely done at the transmitter). Note that matched filtering
introduces a correlation between the successive noise samples. If
desired the binary input bits can be precoded into the bits $b_{i}$
according to the modulo-2 addition
\begin{equation}
b_{i}=\widehat{a}_{i}\oplus b_{i-1}\comma\label{equation3}
\end{equation}
where $\widehat{a}_{i}$ is the complement of $a_{i}$. But neither
matched filtering nor precoding have a significant advantage for \viterbi
decoding.

The states of the system $s_{1}$ and $s_{2}$ are (in symmetrical
form) the bits $-1$ and $+1$ of the binary sequence to be coded.
The state-transitions are given by the corresponding duobinary symbol
(see Table~\ref{table1}). 
\begin{table}[tbph]
\begin{centering}
\begin{tabular}{|c|c|c|}
\hline 
Transition arriving & Possible sequences of bits $a_{i}$ & Duobinary symbol\tabularnewline
\hline 
\hline 
$-1$ & $-1\,\,-1$ & $-1$\tabularnewline
\hline 
 & $+1\,\,-1$ & $0$\tabularnewline
\hline 
\hline 
$+1$ & $-1\,\,+1$ & $0$\tabularnewline
\hline 
 & $+1\,\,+1$ & $+1$\tabularnewline
\hline 
\end{tabular}
\par\end{centering}
\caption{Possible state-transitions. \label{table1}}
\end{table}

If $x$ is the decoder input level without noise and $y$ the same
level with noise, the metric adopted (which is optimal) is the quadratic
difference between $x$ and $y$, expressed as 
\begin{equation}
\delta=(x-y)^{2}=x^{2}-2xy+y^{2}\point\label{equation4}
\end{equation}
Let us note $\gamma_{1,\,p-1}$ and $\gamma_{2,\,p-1}$ the lengths
of the survivors of order $p-1$ and $\lambda_{1,\,p}$ and $\lambda_{2,\,p}$
the lengths of the possible paths arriving at nodes $-1$ and $+1$
at order $p$. These paths are formed by adding the length of the
last branch to the survivors of order $p-1$. This is illustrated
in Table~\ref{table2}. 
\begin{table}[tbph]
\begin{centering}
\begin{tabular}{|c|c|c|c|}
\hline 
Transition & Duobinary symbol & Metric of the last branch & Length of the paths\tabularnewline
\hline 
\hline 
$-1\,\,-1$ & $-1$ & $1+2y+y^{2}$ & $\lambda_{1,\,p}=\gamma_{1,\,p-1}+1+2y+y^{2}$\tabularnewline
\hline 
$+1\,\,-1$ & $0$ & $y^{2}$ & $\lambda_{1,\,p}=\gamma_{2,\,p-1}+y^{2}$\tabularnewline
\hline 
$-1\,\,+1$ & $0$ & $y^{2}$ & $\lambda_{2,\,p}=\gamma_{1,\,p-1}+y^{2}$\tabularnewline
\hline 
$+1\,\,+1$ & $+1$ & $1-2y+y^{2}$ & $\lambda_{2,\,p}=\gamma_{2,\,p-1}+1-2y+y^{2}$\tabularnewline
\hline 
\end{tabular}
\par\end{centering}
\caption{Length of the possible paths at order $p$.\label{table2}}
\end{table}

The first test is to search which of the two values of $\lambda_{1,\,p}$
and $\lambda_{2,\,p}$ is the smallest one. For a transition arriving
at state $-1$, this test can be written as 
\begin{equation}
\gamma_{1,\,p-1}-\gamma_{2,\,p-1}\lessgtr-1-2y\point\label{equation5}
\end{equation}
Note that, in this expression, the symbol $\lessgtr$ denotes the
test ``$<\,\text{OR}\,>$''. If the response to this test is ``$<$''
the shortest path corresponds to the transition $-1\,\,-1$ and the
survivor $S_{1}$ of order $p$ prolongs the survivor $S_{1}$ of
order $p-1$. On the other hand if the response is ``$>$'' the
survivor $S_{1}$ of order $p$ prolongs the survivor $S_{2}$ of
order $p-1$. A similar test is made on the transitions arriving at
state $+1$, and so on.

\subsection{Other partial response codes}

Apart from the duobinary or Class~1 code, there are four other non-extended
partial response codes defined as follows by their polynomial $P(D)$~\cite{Spilker1977Digital}: 
\begin{itemize}
\item Class 2: $P(D)=1+2D+D^{2}$,
\item Class 3: $P(D)=2+D-D^{2}$,
\item Class 4: $P(D)=1-D$,
\item Class 5: $P(D)=1+2D^{2}-D^{4}$.
\end{itemize}
Note that the Class~3 code may be of some interest for our application
because of its good resistance to echoes in AM/VSB modulation~\cite{Mertens1989Echo}.

Extended partial response codes have a polynomial of the form $P(D)=(1-D)^{m}(1+D)^{n}$,
duobinary coding being a sub-case for which $m=0$ and $n=1$. These
codes have been extensively studied in particular for application
like data magnetic recording and disk storage, where as already mentioned,
the error-rate should be very low (\ie a high signal to noise ratio).

For example reference~\cite{Filho1999AMultilevel} describes techniques
for the construction of ``good'' codes (with $n=1$) by matching
the trellis to the channel and using a ``pre-coder'' which can be
a punctured convolutional code. Reference~\cite{Leung2001Reduced}
is an attempt to simplify the\noun{ }\viterbi decoder by deleting
the less probable branches in the trellis. Reference~\cite{Altekar1999Error}
is a systematic search of state-sequences for which the squared Euclidean
distance from the decoded sequence has a given value. This paper will
be analyzed later (see Section~\ref{subsec:Upper-bound-of}) where
we give a theoretical upper bound for the final error-rate with duobinary
coding.

The paper of Vityaev and Siegel~\cite{Vityaev1998OnViterbi} gives,
by solving a series of problems in linear programming, upper and lower
bounds of the metric difference between two paths of same length in
the trellis but arriving at two different states. This consideration
is essential for the practical implementation of the \viterbi decoder
(see Section~\ref{subsec:Implementation-of-the-Viterbi}). Despite
of its practical importance this aspect is not dealt with in this
paper where we are not considering any implementation of a duobinary
\viterbi decoder in the form of integrated circuits, some of them
having already been developed by the industry.

\subsection{Implementation of the \viterbi decoder\label{subsec:Implementation-of-the-Viterbi}}

The results given in this paper were always obtained by software (\ie
a number of computer programs specially written). In hardware implementations,
the received signal as well as the metrics and the survivors path
should be, in discrete form, represented by binary numbers with a
sufficient number of bits and a small enough quantizing interval,
while assuring a sufficiently high speed and hardware simplicity.

Reference~\cite{UER-Doc-GTR5-101} gives an example of a CMOS \nomenclature{CMOS}{Complementary Metal-Oxide Semiconductor}
implementation developed at the CCETT (France) of a \viterbi decoder
for duobinary signals, with a discussion of the choices made for quantization.
Risks of overflow also need to be addressed carefully. Bounds of metrics,
as derived by Vityaev and Siegel~\cite{Vityaev1998OnViterbi}, are
useful to avoid overflows in the registers which could otherwise produce
long error bursts.

The paper of\noun{ }Chang~\cite{Chang2003AnEfficient} presents another
design of \viterbi decoder based on in-place state metric update
and hybrid survivor path management. It describes the three main functional
blocks of the decoder: a Branch Metric Unit which receives the noisy
symbols and computes the corresponding branch metric, the Add-Compare-Select
(ACS\nomenclature{ACS}{Add-Compare-Select}) unit which updates the
metrics for all states by proper comparison (\ie what in this paper
we call the \emph{tests} made by the decoder) and Trace-Back Unit
which searches the best survivor path on the basis of accumulated
decisions. The design of these units result in an efficient architecture
for a \viterbi decoder specially in the cases of trellis with a large
number of states and a large value of the constraint length, for example
for decoding convolutional codes or concatenations of two codes.

The paper of Zand and Johns~\cite{Zand2002High} describes an analog
CMOS \viterbi detector for use on a 4-PAM\nomenclature{PAM}{Phase-Amplitde modulation}
duobinary signaling. The chip is an implementation of a reduced state
sequence with pipelining and parallel processing and the application
studied is optical links.

It is also known that implementations of \viterbi decoders have been
developed by industry, even if the results have not been published
as generally available papers.

\subsection{Some properties of the duobinary code}

\subsubsection{Duobinary eye diagram\label{subsec:Duobinary-eye-diagram-ex4}}

\begin{figure}[tbph]
\begin{centering}
\includegraphics[width=13cm]{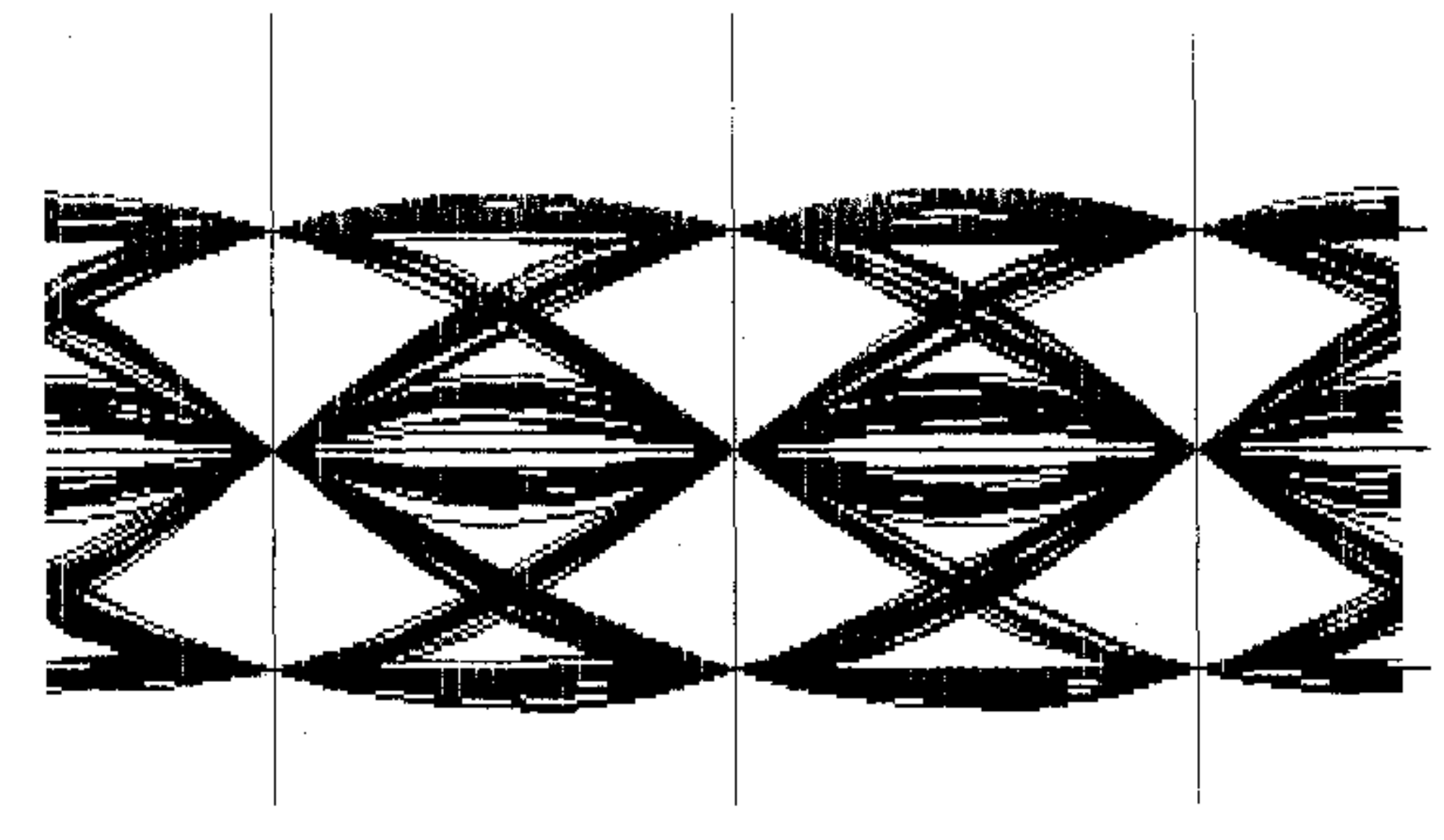}
\par\end{centering}
\caption{Duobinary eye diagram (by simulation).\label{figure2}}
\end{figure}

In order to compute the error-rate in presence of a synchronization
error or of an echo, it is necessary to know the decoder input level
at any time and not only at the nominal sampling instants. This can
be given by the eye diagram. A typical duobinary eye diagram, obtained
by simulation with a random sequence, is given in Figure~\ref{figure2}.
It is seen that there is no inter-symbol interference at the nominal
sampling instants $kT$, where $T$ denotes the duration of one bit.

\begin{figure}[tbph]
\begin{centering}
\includegraphics[width=13cm]{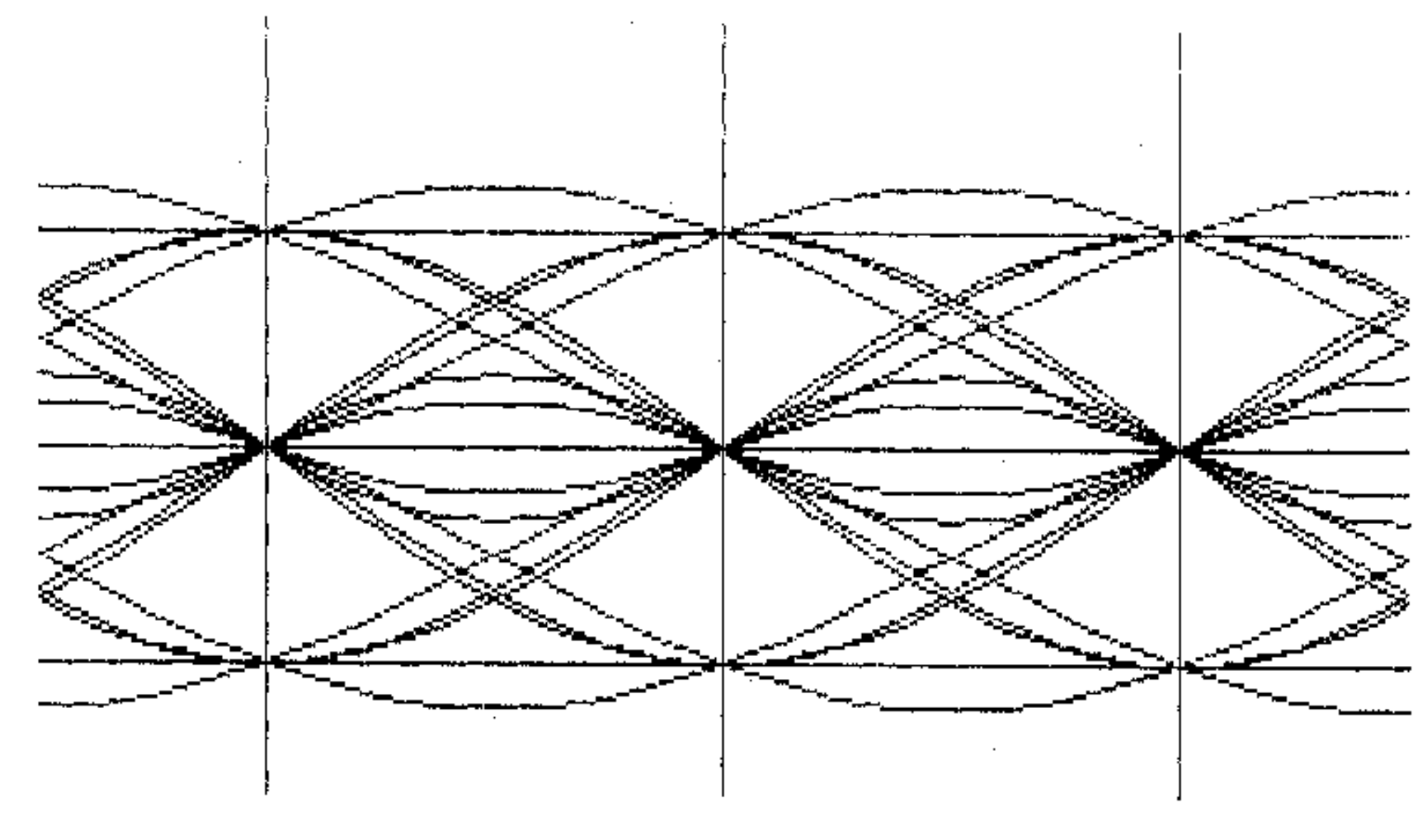}
\par\end{centering}
\caption{Reconstructed duobinary eye diagram (first approximation).\label{figure3}}
\end{figure}

A first approximation of the level of the diagram at any time $t$
is obtained by superposing a large number of impulse responses of
the duobinary filter, shifted by $n$ bit periods, including all possible
binary sequences of $+1$ and $-1$ up to length 4, and by developing
the result in Fourier series. It is found that the level can take
twenty analytical values with equal probabilities. For illustration,
only the first four values are given hereafter, with a $\pm$ sign:
\begin{equation}
x_{1}(t)=\pm1\comma\label{equation6}
\end{equation}
\begin{equation}
x_{2}(t)=\pm\left(\frac{1}{3}+\frac{2}{3}\,\cos\left(\frac{2\pi t}{3T}\right)\right)\point\label{equation7}
\end{equation}
The eye diagram reconstructed by superposing these twenty forms is
shown in Figure~\ref{figure3}.

\begin{figure}[tbph]
\begin{centering}
\includegraphics[width=13cm]{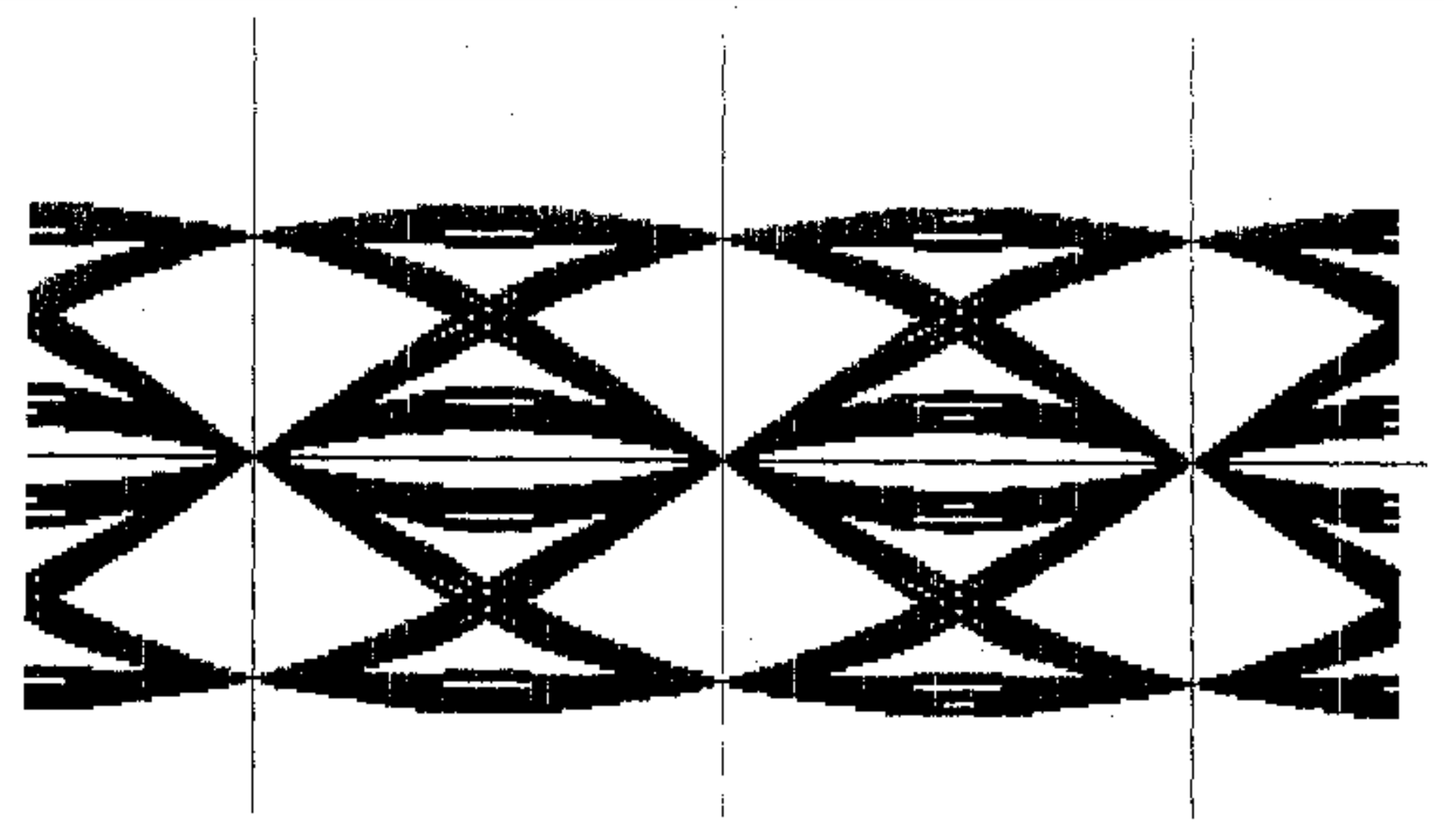}
\par\end{centering}
\caption{Reconstructed duobinary eye diagram (second approximation).\label{figure4}}
\end{figure}

A second (and better) approximation is obtained by extending this
procedure to sequences of $-1$ and $+1$ up to length 9. Development
in Fourier series was not made for this case because the number of
possible analytical forms becomes too large. Instead, all the possible
levels of the eye diagram were stored in a computer file with a sampling
frequency of 16 times the bit frequency. This second approximation
is given in Figure~\ref{figure4}.

\begin{figure}[tbph]
\begin{centering}
\includegraphics[width=13cm]{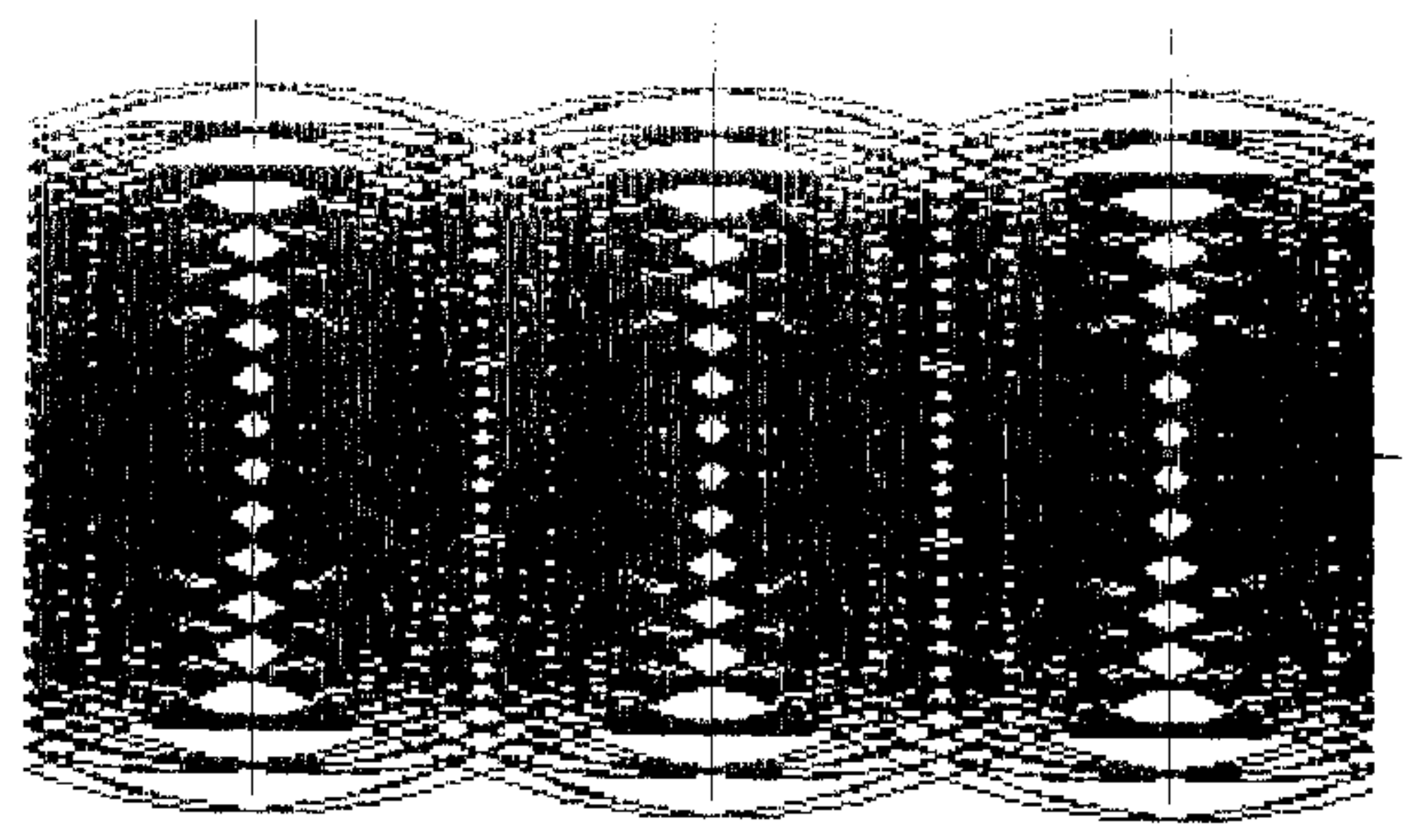}
\par\end{centering}
\caption{Eye diagram of a duobinary signal in quadrature.\label{figure5}}
\end{figure}

For the case of AM/SSB modulation, the same method was used to obtain
the eye diagram of the duobinary signal in quadrature by filtering
all duobinary sequences obtained after the duobinary filter by the
quadrature filter of transfer function
\begin{equation}
G_{q}(f)=e^{-j\frac{\pi}{2}}\,sign(f)\point\label{equation8}
\end{equation}
All the levels were again stored in a computer file and the corresponding
eye diagram is given in Figure~\ref{figure5}.

\subsubsection{Correlation of the duobinary signal}

The correlation coefficients between duobinary symbols separated by
$jT$ were computed for the following cases:
\begin{itemize}
\item $D$: autocorrelation of a duobinary signal,
\item $DQ$ : autocorrelation of a duobinary signal in quadrature,
\item $I$: intercorrelation between $D$ and $DQ$,
\end{itemize}
with the same values for negative separation $j$, except a change
of sign for the intercorrelation $I$. Results are drawn in Table~\ref{table3}.
\begin{table}[tbph]
\begin{centering}
\begin{tabular}{|c|c|c|c|}
\hline 
Separation $j$ & $D$ & $DQ$ & $I$\tabularnewline
\hline 
\hline 
6 & $\left|<10^{-2}\right|$ & $-1.8\times10^{-2}$ & $1.3\times10^{-2}$\tabularnewline
\hline 
5 & $\left|<10^{-2}\right|$ & $-8.0\times10^{-2}$ & $\left|<10^{-2}\right|$\tabularnewline
\hline 
4 & $\left|<10^{-2}\right|$ & $-0.135$ & $-4.7\times10^{-2}$\tabularnewline
\hline 
3 & $\left|<10^{-2}\right|$ & $-0.109$ & $-0.105$\tabularnewline
\hline 
2 & $\left|<10^{-2}\right|$ & $-6.9\times10^{-2}$ & $-0.211$\tabularnewline
\hline 
1 & $0.25$ & $0.189$ & $-0.312$\tabularnewline
\hline 
0 & $0.5$ & $0.440$ & $\left|<10^{-2}\right|$\tabularnewline
\hline 
\end{tabular}
\par\end{centering}
\caption{Duobinary correlation coefficients.\label{table3}}
\end{table}

These values are considered in Section~\ref{sec:Interferences} for
the case of a correlated interference due to another duobinary sequence. 

\subsubsection{Probability of the duobinary symbols}

If the binary input data $-1$ and $1$ have the same probability,
the probabilities of the duobinary symbols $-1,\,0,\,1$ are respectively
$1/4$, $1/2$ and $1/4$.

\section{Theoretical framework for duobinary coding and the \viterbi decoding
\label{sec:Duobinary-theoretical-framework}}

A detailed description of the \viterbi decoding is given in references~\cite{Bossert1999Channel,Forney1973TheViterbi,Lender1963TheDuobinary,Viterbi1971Convolutional}.
The length of the possible paths at order $p$ in the trellis was
already given in Table~\ref{table2} above.

\subsection{Formulation of the tests made by the \viterbi decoder\label{subsec:Formulation-of-the-test}}

In order to construct the survivors at order $p$, the decoder looks
for the shortest path between the two values of $\lambda_{1,\,p}$
and $\lambda_{2,\,p}$. If the decoder keeps in memory the \emph{renormalized
metric} $\mu_{p}$, which is the difference in length between the
two survivors, we have 
\begin{equation}
\mu_{p-1}=\gamma_{1,\,p-1}-\gamma_{2,\,p-1}\comma\label{equation9}
\end{equation}
\begin{equation}
\mu_{p}=\gamma_{1,\,p}-\gamma_{2,\,p}\point\label{equation10}
\end{equation}
 The tests on the path lengths are then 
\begin{equation}
\mu_{p-1}+2y\lessgtr-1\comma\label{equation11}
\end{equation}
\begin{equation}
\mu_{p-1}+2y\test1\point\label{equation12}
\end{equation}
 It can be shown that three different situations can occur with the
following consequences: 
\begin{enumerate}
\item If
\begin{equation}
\mu_{p-1}+2y<-1\label{equation13}
\end{equation}
then
\begin{equation}
\mu_{p}=1+2y\point\label{equation14}
\end{equation}
The decoder recognizes the duobinary symbol $-1$ and the survivors
$S_{1}$ and $S_{2}$ of order $p$ prolong the survivor $S_{1}$
at order $p-1$; in other words the survivor $S_{1}$ is copied into
the new survivors $S_{1}$ and $S_{2}$.
\item If
\begin{equation}
-1\leq\mu_{p-1}+2y\leq1\label{equation15}
\end{equation}
then
\begin{equation}
\mu_{p}=-\mu_{p-1}\point\label{equation16}
\end{equation}
The decoder recognizes the duobinary symbol 0 and the survivor $S_{1}$
at order $p$ prolongs the survivor $S_{2}$ at order $p-1$, while
the survivor $S_{2}$ at order $p$ prolongs the survivor $S_{1}$,
so that $S_{1}$ is copied in the new survivor $S_{2}$ and $S_{2}$
is copied in the new survivor $S_{1}$. In other words, survivors
are \emph{crossed}. 
\item If
\begin{equation}
\mu_{p-1}+2y>1\label{equation17}
\end{equation}
then
\begin{equation}
\mu_{p}=1-2y\point\label{equation18}
\end{equation}
The decoder recognizes the duobinary symbol $1$ and the survivors
$S_{1}$ and $S_{2}$ at order $p$ prolong the survivor $S_{2}$
of order $p-1$; in other words $S_{2}$ is copied into the new survivors
$S_{1}$ and $S_{2}$.
\end{enumerate}
After these tests $-1$ is added to $S_{1}$ and $1$ is added to
$S_{2}$ and the two survivors are truncated and shifted by one position.
Since the two survivors are converging, decoding of the input bit
$a_{i}$ can be made after a limited number of tests, the \emph{decoding
constraint length} being typically 20 to 30 bits.

\subsection{Alternative formulation of the tests}

In the following, we consider additive noise. We may therefore rewrite
$y_{p}$ as 
\begin{equation}
y_{p}=d_{p}+n_{p}\point\label{equation19}
\end{equation}
Assume that, at order $p$, 
\begin{equation}
\mu_{p}+2y_{p}=\mu_{p}+2(d_{p}+n_{p})<-1\point
\end{equation}
The next value for the normalized metric is 
\begin{equation}
\mu_{p+1}=1+2y_{p}=1+2(d_{p}+n_{p})\comma
\end{equation}
so that the new test is made on the quantity
\begin{equation}
\mu_{p+1}+2y_{p+1}=1+2(d_{p}+n_{p})+2(d_{p+1}+n_{p+1})\point\label{equation22}
\end{equation}
 If this test is again giving the answer $<-1$, then
\begin{equation}
\mu_{p+2}=1+2y_{p+1}\comma
\end{equation}
and so on.

If the test (\ref{equation22}) gives the answer $>1$, then
\begin{equation}
\mu_{p+2}=1-2y_{p+1}=1-2(d_{p+1}+n_{p+1})\point
\end{equation}
If finally the quantity (\ref{equation22}) is between $-1$ and $1$,
\begin{equation}
\mu_{p+2}=-\mu_{p+1}=-1-2(d_{p}+n_{p})\comma
\end{equation}
and the test $p+2$ is made on the quantity
\begin{equation}
\mu_{p+2}+2y_{p+2}=-1-2(d_{p}+n_{p})+2(d_{p+2}+n_{p+2})\comma
\end{equation}
where the duobinary symbols and the noise samples are $d_{p}$ and
$d_{p+2}$, and $n_{p}$ and $n_{p+2}$, respectively. Therefore as
long as the decoder recognizes duobinary symbols 0, the test involves
the transmitted symbol $d_{p}$ and the noise sample $n_{p}$ corresponding
to the last test which gave the answer $<-1$ or $>1$. 

With a slight change in the notations, the following rules can therefore
be established.
\begin{enumerate}
\item Last test ($i-k$) with answer $>1$ \\
Test $i$ is made on the quantity
\begin{equation}
2(d_{i}+d_{i-k})+2(n_{i}+n_{i-k})-1\,\,\,\,\,\,\,\,\,\,\textrm{with }k\textrm{ odd}\label{equation27}
\end{equation}
\begin{equation}
2(d_{i}-d_{i-k})+2(n_{i}-n_{i-k})+1\,\,\,\,\,\,\,\,\,\,\textrm{with }k\textrm{ even}
\end{equation}
\item Last test ($i-k$) with answer $<-1$ \\
Test $i$ is made on the quantity
\begin{equation}
2(d_{i}+d_{i-k})+2(n_{i}+n_{i-k})+1\,\,\,\,\,\,\,\,\,\,\textrm{with }k\textrm{ odd}
\end{equation}
\begin{equation}
2(d_{i}-d_{i-k})+2(n_{i}-n_{i-k})-1\,\,\,\,\,\,\,\,\,\,\textrm{with }k\textrm{ even}\label{equation30}
\end{equation}
\end{enumerate}

\subsection{Duobinary error probability with low noise}

If there is no noise, the symbol $r_{i}$ recognized by the decoder
is always the transmitted symbol $d_{i}$.

If we consider a very low level of the noise, the duobinary error-rate
is not zero. Suppose $d_{i}=1$ is the transmitted symbol. The test
$i$, as given by (\ref{equation27}) to (\ref{equation30}), provides
a value close to $3$ which is well separated from the limit $1$
and there is no error. If $d_{i}=-1$, test $i$ gives a value close
to $-3$ and again there is no error. If finally $d_{i}=0$, test
$i$ gives a value close to $1$ or to $-1$ and there is therefore
a probability equal to $1/2$ that the decoder recognizes $1$ (or
$-1$) instead of $0$. 

The mean error-rate can be found by computing the error probability
on the last $0$ of the duobinary sequences:
\begin{itemize}
\item $10$ and $-10$ where the probability of the symbols before the last
$0$ is $1/4$,
\item $100$ and $-100$ where the probability of the symbols before the
last $0$ is $1/8$,
\end{itemize}
and so on.

It can be shown that, with the method of numerical integrations described
later in Sections~\ref{subsec:Probability-of-a-recognized} and \ref{subsec:Duobinary-error-rate-10},
the duobinary error-rate with low noise at length $k$ is in fact
$1/k$. By summing the terms $1/k$ for $k=2,\,3,\,4,\,\ldots$, weighted
by the probabilities of the symbols before the last $0$, we see that
the mean duobinary error probability is given by 
\begin{equation}
P_{d}(\textrm{low noise})=\frac{1}{2}\,\frac{1}{4}+\frac{1}{3}\,\frac{1}{8}+\frac{1}{4}\,\frac{1}{16}+\ldots\comma
\end{equation}
\begin{equation}
P_{d}(\textrm{low noise})=\sum_{k=2}^{\infty}\,\frac{1}{k\,2^{k}}=0.193\point
\end{equation}
This value was confirmed by simulation. In the next section, we examine
the effect for any level of noise. 

\subsection{Probability of a recognized sequence $r$ for a transmitted sequence
$d$\label{subsec:Probability-of-a-recognized}}

For any level of the noise, if $d_{i}$ are the transmitted duobinary
symbols and $r_{i}$ the symbols decoded (correctly or not), the tests
(\ref{equation27}) to (\ref{equation30}) where $d_{i}$ and $r_{i}$
are replaced by their values, take the form
\begin{equation}
n_{i}<a\mp n_{i-k}\comma
\end{equation}
\begin{equation}
n_{i}>b\mp n_{i-k}\comma
\end{equation}
the limits $a$ and $b$ depending on $d_{i}$, $d_{i-k}$, $r_{i}$
and $r_{i-k}$.

As an example, let us consider the transmitted sequence
\begin{equation}
d_{i}=1000\comma
\end{equation}
and the recognized sequence
\begin{equation}
r_{i}=1(-1)00\point
\end{equation}
(Note that we will always start the sequences $d_{i}$ and $r_{i}$
with $1$ or $-1$ in order to start the calculation with a test $i-k$
of known value). 

The tests made by the decoder can be written in this case
\begin{equation}
n_{1}<-1-n_{2}\comma\label{equation37}
\end{equation}
\begin{equation}
n_{3}<-n_{2}\,\,\,\,\,\textrm{and}\,\,\,\,\,n_{3}>-1-n_{2}\comma
\end{equation}
\begin{equation}
n_{4}<1+n_{2}\,\,\,\,\,\textrm{and}\,\,\,\,\,n_{4}>n_{2}\point\label{equation39}
\end{equation}
The problem is to compute the probability of a system of simultaneous
inequalities like (\ref{equation37}) to (\ref{equation39}), where
the $n_{i}$ are independent samples of Gaussian noise with variance
$\sigma^{2}$. However note that the same noise samples (in this case
$n_{2}$) can appear in several inequalities.

Let $g(n)$ be the Gaussian probability density distribution
\begin{equation}
g(n_{i})=\frac{1}{\sigma\sqrt{2\pi}}\,e^{-\frac{n_{i}^{2}}{2\sigma^{2}}}\point
\end{equation}
The probability $P_{r}$ of (\ref{equation37}) to (\ref{equation39})
is given by
\begin{equation}
P_{r}=\int_{-\infty}^{+\infty}g(n_{2})\,\int_{-\infty}^{-1-n_{2}}g(n_{1})\,\int_{-1-n_{2}}^{-n_{2}}g(n_{3})\,\int_{n_{2}}^{1+n_{2}}g(n_{4})\,dn_{2}\,dn_{1}\,dn_{3}\,dn_{4}\comma\label{equation41}
\end{equation}
which for a sequence of length $L$ is, in general, an $L$-uple integral.

However, if a particular noise sample (in this case $n_{2}$) is common
to several successive integrals, the calculation is simplified by
introducing the $Q$ function
\begin{equation}
Q(x)=\frac{1}{\sqrt{2\pi}}\int_{x}^{\infty}e^{-\frac{u^{2}}{2}}du\comma
\end{equation}
and (\ref{equation41}) reduces to the simple integral
\[
P_{r}=\int_{-\infty}^{+\infty}\left[1-Q\left(\frac{-1-n_{2}}{\sigma}\right)\right]\left\{ \left[1-Q\left(\frac{-n_{2}}{\sigma}\right)\right]-\left[1-Q\left(\frac{-1-n_{2}}{\sigma}\right)\right]\right\} 
\]
\begin{equation}
\left\{ \left[1-Q\left(\frac{1+n_{2}}{\sigma}\right)\right]-\left[1-Q\left(\frac{n_{2}}{\sigma}\right)\right]\right\} \,dn_{2}\point
\end{equation}
Such a simplification is not always possible, and there are cases
of length $L=4$ where $P_{r}$ is given by a double integral, of
length 5 by a triple integral, and so on.

\subsection{Duobinary error-rate\label{subsec:Duobinary-error-rate-10}}

The duobinary error-rate due to an incorrect recognition of $0$ is
obtained by computing the error probability on the last $0$ of the
following duobinary sequences (of which the total probability of the
symbols before the last $0$ is $1/2+1/4+1/8+\,\ldots=1$): 
\begin{itemize}
\item $L=2:$ ~~~$10$ and $-10$,
\item $L=3:$ ~~~$100$ and $-100$,
\item $L=4:$ ~~~$1000$ and $-1000$,
\end{itemize}
and so on. Sequences such as $110$ or $-1-10$ should be excluded
because we always start with a test on $1$ or $-1$ followed by zeros.

The total error probability on $0$ is then
\begin{equation}
P_{d0}=\frac{1}{2}\,p_{02}+\frac{1}{4}\,p_{03}+\frac{1}{8}\,p_{04}+\ldots\comma\label{equation44}
\end{equation}
which is a sum weighted by the probabilities of the corresponding
sequences.

Similarly, the duobinary error-rate due to a false recognition of
$1$ (or $-1$) is obtained by computing the error probability on
the last $1$ (or $-1$) of the following duobinary sequences: 
\begin{itemize}
\item $L=2:$~~~$11$,
\item $L=3$:~~~$-101$,
\item $L=4$:~~~$1001$,
\end{itemize}
and so on.

The corresponding error-rate is
\begin{equation}
P_{d1}=P_{d-1}=\frac{1}{2}\,p_{12}+\frac{1}{4}\,p_{13}+\frac{1}{8}\,p_{14}+\ldots\label{equation45}
\end{equation}
The error probabilities on $0$ and $1$ of all these sequences, up
to length $L=5$, were computed by the expressions of the previous
Sections, using the software ``Mathematica'' for integration.

For the length $L=6,\,7,\,8$, the theoretical calculation involves
quadruple, quintuple and sextuple integrals, and the computation time
at each step is multiplied by a significant factor. A data base of
all possible duobinary errors was therefore filled in by simulation,
not of the \viterbi  decoding itself, but on the tests on the noise
samples expressed by a system of inequalities such as (\ref{equation37})
to (\ref{equation39}).

Because the terms of the series (\ref{equation44}) and (\ref{equation45})
are decreasing, the contribution of the lengths larger than 8 is negligible
and the dominant terms of the final results were computed analytically. 

\begin{figure}[htbp]
\begin{centering}
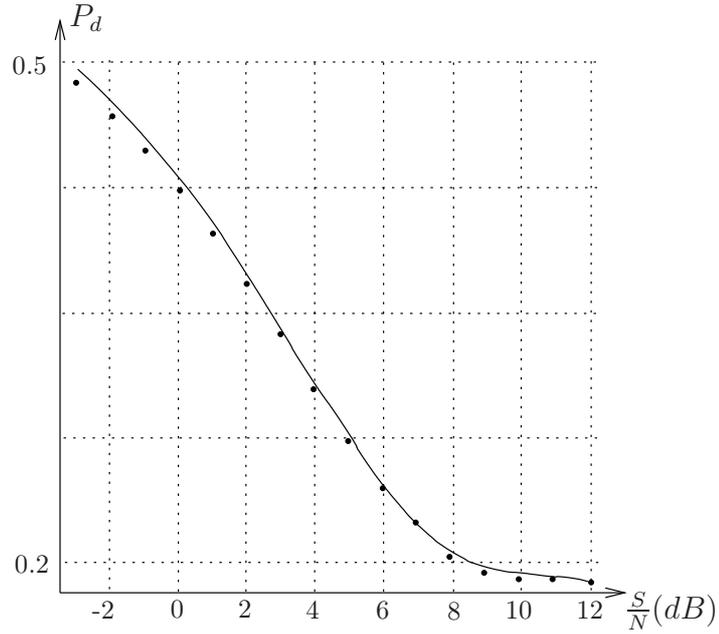
\par\end{centering}
\caption{Duobinary error-rate as a function of $S/N$: theoretical curve and
points of simulation.\label{figure6}}
\end{figure}

Since the error probabilities on $1$ and $-1$ are equal (due to
the symmetry of the code), the final mean duobinary error-rate is
\begin{equation}
P_{d}=\frac{1}{2}\,(P_{d0}+P_{d1})\point
\end{equation}
This theoretical result and values obtained by simulation are drawn
on Figure~\ref{figure6} as a function of the signal-to-noise ratio
$S/N$ (in baseband with unmatched filtering). As shown by this Figure,
the present theory is well confirmed by simulation.

\subsection{Error-rate on the survivors}

Let us call a \emph{simple} \emph{error} an error at one position
on one of the survivors and a \emph{double} \emph{error} an error
which affects both survivors at the same position.

The computation of the error-rate on the survivors is complex but
we can use the following rules that we have established: 
\begin{itemize}
\item If the duobinary transmitted symbol $0$ is recognized as $-1$, there
are $n+1$ consecutive simple errors on the survivor $S_{1}$, $n$
being the number of times that a symbol $0$ is recognized correctly
before the duobinary error.
\item If a transmitted $0$ is recognized as $1$, there are similarly $n+1$
consecutive simple errors on $S_{2}$.
\item If a transmitted $-1$ is recognized as $0$, there are $n+1$ consecutive
simple errors on $S_{1}$.
\item If a transmitted $1$ is recognized as $0$, there are $n+1$ consecutive
simple errors on $S_{2}$.
\item If a transmitted $1$ (or $-1$) is recognized as $-1$ (or $1$),
there are $n+1$ consecutive double errors on $S_{1}$ and $S_{2}$.
\end{itemize}
If $P_{0}=1/2$ is the probability of symbol $0$ and $P_{d0}$ the
duobinary error probability on $0$, the probability of one simple
error on $S_{1}$ (or $S_{2}$) due to a duobinary error on $0$ is
given by
\begin{equation}
P_{s,\,1}(0)=[1-P_{0}(1-P_{d0})]P_{0}P_{d0}\comma
\end{equation}
the term within brackets representing the probability of having anything
other than a $0$ correctly recognized before the duobinary error.

\begin{figure}[tbph]
\begin{centering}
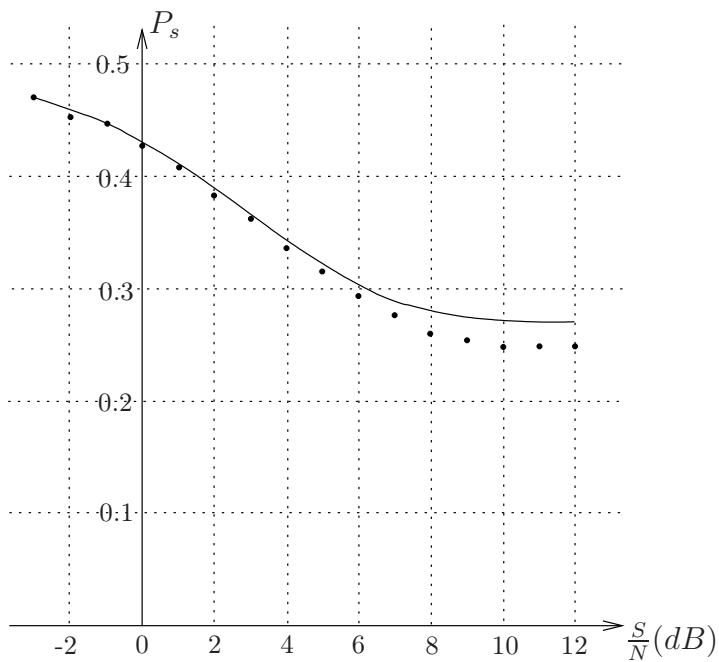
\par\end{centering}
\caption{Simple error-rate on the survivors as a function of $S/N$: theoretical
curve and points of simulation.\label{figure7}}
\end{figure}

\begin{figure}[tbph]
\begin{centering}
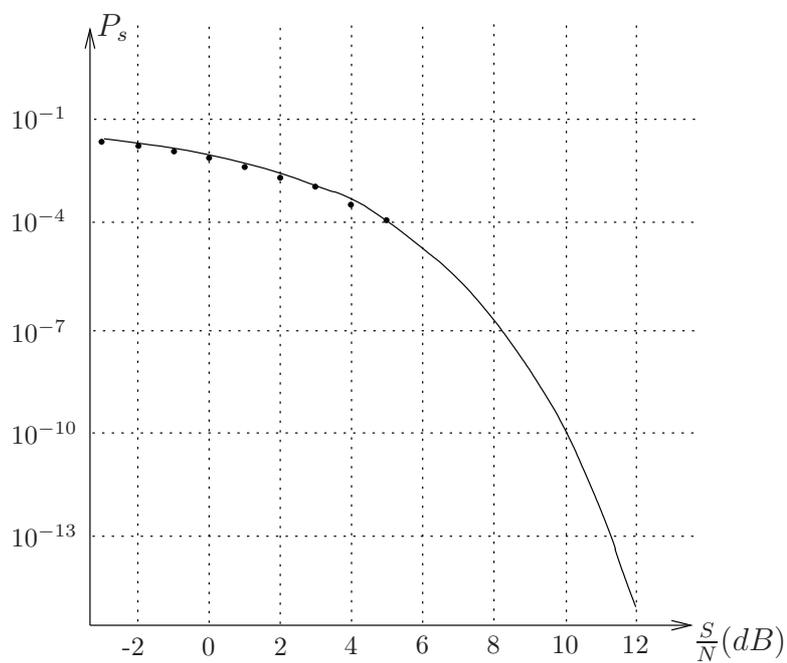
\par\end{centering}
\caption{Double error-rate on the survivors as a function of $S/N$: theoretical
curve and points of simulation. \label{figure8}}
\end{figure}

The probability to have $k$ consecutive simple errors is then
\begin{equation}
P_{s,\,k}(0)=[1-P_{0}(1-P_{d0})]P_{0}^{k-1}(1-P_{d0})^{k-1}P_{0}P_{d0}\point
\end{equation}
 If $P_{1}=1/4$ is the probability of symbol $1$ and $P_{d10}$
the probability of a duobinary error on a symbol $1$ recognized as
$0$, the probabilities of a single simple error on $S_{1}$ or $S_{2}$
or of $k$ consecutive simple errors are
\begin{equation}
P_{s,\,1}(1)=[1-P_{0}(1-P_{d0})]P_{1}P_{d10}\comma
\end{equation}
\begin{equation}
P_{s,\,k}(1)=[1-P_{0}(1-P_{d0})]P_{0}^{k-1}P_{1}(1-P_{d0})^{k-1}P_{d10}\comma
\end{equation}
and for the double errors we have
\begin{equation}
P_{sd,\,1}(1)=[1-P_{0}(1-P_{d0})]P_{1}P_{d1-1}\comma
\end{equation}
\begin{equation}
P_{sd,\,k}(1)=[1-P_{0}(1-P_{d0})]P_{0}^{k-1}P_{1}(1-P_{d0})^{k-1}P_{d1-1}\comma
\end{equation}
where $P_{d1-1}$ is the probability that $1$ (or $-1$) is recognized
as $-1$ (or $1$). The mean rate of simple and double errors on the
survivors were computed with these expressions and compared to values
obtained by simulation. The results are given in Figures~\ref{figure7}
and~\ref{figure8}. 

\subsection{Evolution of the errors on the survivors}

The necessary and sufficient condition for a binary error to occur
is a double error on the survivors, because such an error will never
be eliminated if $S_{1}$ or $S_{2}$ are \emph{recopied} or \emph{crossed}
(see Section~\ref{subsec:Formulation-of-the-test}). Fortunately
during the process of survivors convergence, most of the simple errors
will be eliminated, while some of these simple errors are transformed
into double errors. Indeed, let us suppose that a single simple error
occurs at step $i$ on $S_{1}$ and that, after the error, the decoder
recognizes $j-1$ symbols $0$. Then the survivors are crossed $j-1$
times. At step $i+1$ the error is transported at $S_{2}$, at step
$i+2$ again at $S_{1}$, and so on. If now at step $i+j$, the decoder
recognizes a symbol $-1$, the survivor $S_{1}$ of step $i+j-1$
is copied into $S_{1}$ and $S_{2}$. If $j$ is odd, then the simple
error becomes a double error, while the error is eliminated if $j$
is even.

Generalization of this reasoning gives the following rules for the
conditions of transformation of a simple error into a double error
and thus into a binary error

\subsubsection{Normal rules}
\begin{enumerate}
\item Simple error at step $i$ on $S_{1}$ becomes a double error if

\begin{itemize}
\item test $i+j$ gives $<-1$ with $j$ odd.
\item test $i+j$ gives $>1$ with $j$ even.
\end{itemize}
\item Simple error at step $i$ on $S_{2}$ becomes a double error if

\begin{itemize}
\item test $i+j$ gives $<-1$ with $j$ even.
\item test $i+j$ gives $>1$ with $j$ odd.
\end{itemize}
\end{enumerate}

\subsubsection{Additional rules}

The rules given above are only valid if there are no new errors on
the survivors between test $i$ (where there is an initial error)
and test $i+j$ (where the initial error is eliminated or transformed
into a binary error). This is, in fact, the case when the binary error-rate
is low.

If the signal is affected by a strong noise or distortion, the probability
of new errors between $i$ and $i+j$ is no longer negligible and
the behavior of the decoder becomes more complex. Some additional
rules for the evolution of the survivors errors must then be introduced.
The most frequent of them are given below. 
\begin{enumerate}
\item \emph{Simple errors at $i$ followed by simple errors at $i+m$ with
$m<j$}. \\
Let $IS1(i)$ and $IS2(i)$ be the numbers of simple errors occurring
on survivors $S_{1}$ and $S_{2}$ at test $i$. We have: \\
\\
\begin{tabular}{|c|c|}
\hline 
If & Then everything is as if\tabularnewline
\hline 
\hline 
$IS1(i)>0,IS2(i+m)>0,r(i)=-1,m\textrm{ odd }<j$ & $IS1(i)=0$\tabularnewline
\hline 
$IS2(i)>0,IS1(i+m)>0,r(i)=1,m\textrm{ odd }<j$ & $IS2(i)=0$\tabularnewline
\hline 
$IS1(i)>0,IS1(i+m)>0,r(i)=-1,m\textrm{ even }<j$ & $IS1(i)=0$\tabularnewline
\hline 
$IS2(i)>0,IS2(i+m)>0,r(i)=1,m\textrm{ even }<j$ & $IS2(i)=0$\tabularnewline
\hline 
\end{tabular}
\item \emph{Simple errors at $i$ followed by double errors at $i+m$ with
$m=j$.} \\
Let in addition $ID(i)$ be the number of double errors on the survivors
at test $i$. We have: \\
\\
\begin{tabular}{|c|c|}
\hline 
If & Then everything is as if\tabularnewline
\hline 
\hline 
$IS1(i)>0,ID(i+m)>0,$ & $IS1(i)=0$\tabularnewline
$r(i)=-1,r(i+m)=-1,m\textrm{ odd }=j$ & \tabularnewline
\hline 
$IS2(i)>0,ID(i+m)>0,$ & $IS2(i)=0$\tabularnewline
$r(i)=1,r(i+m)=1,m\textrm{ odd }=j$ & \tabularnewline
\hline 
$IS1(i)>0,ID(i+m)>0,$ & $IS1(i)=0$\tabularnewline
$r(i)=-1,r(i+m)=1,m\textrm{ even }=j$ & \tabularnewline
\hline 
$IS2(i)>0,ID(i+m)>0,$ & $IS2(i)=0$\tabularnewline
$r(i)=1,r(i+m)=-1,m\textrm{ even }=j$ & \tabularnewline
\hline 
\end{tabular}
\item \emph{Non transformed errors: simple errors at $i$ followed by simple
errors at $i+m$ with $m=j$.}\\
Let now $IS1N(i)$ and $IS2N(i)$ be the numbers of simple errors
on survivors $S_{1}$ and $S_{2}$ that would be transformed in double
errors at test $j$ according to the normal rules but that are in
fact not transformed. Then\\
\\
\begin{tabular}{|c|c|}
\hline 
If & Then\tabularnewline
\hline 
\hline 
$IS1(i)>0,IS1(i+m)>0,$ & $IS1N(i)=IS1(i)$\tabularnewline
$r(i)=-1,r(i+m)=-1,m\textrm{ odd }=j$ & \tabularnewline
\hline 
$IS2(i)>0,IS2(i+m)>0,$ & $IS1N(i)=IS1(i)$\tabularnewline
$r(i)=1,r(i+m)=1,m\textrm{ odd }=j$ & \tabularnewline
\hline 
$IS1(i)>0,IS2(i+m)>0,$ & $IS1N(i)=IS1(i)$\tabularnewline
$r(i)=-1,r(i+m)=1,m\textrm{ even }=j$ & \tabularnewline
\hline 
$IS2(i)>0,IS1(i+m)>0,$ & $IS1N(i)=IS1(i)$\tabularnewline
$r(i)=1,r(i+m)=-1,m\textrm{ even }=j$ & \tabularnewline
\hline 
\end{tabular}
\item \emph{Evolution of non transformed errors.}\\
Let $i+j+j_{1}$ be the first test after $i+j$ where the symbol recognized
by the decoder is $-1$ or $1$. Then if $IB(i)$ is the number of
\emph{binary} errors due to duobinary errors at test $i$:\\
\\
\begin{tabular}{|c|c|}
\hline 
If & Then also\tabularnewline
\hline 
\hline 
$IS1N(i)>0,j\textrm{ even},j_{1}\textrm{ even},r(i+j+j_{1})=1$ & $IB(i)=IS1N(i)$\tabularnewline
\hline 
$IS1N(i)>0,j\textrm{ even},j_{1}\textrm{ odd},r(i+j+j_{1})=-1$ & $IB(i)=IS1N(i)$\tabularnewline
\hline 
$IS1N(i)>0,j\textrm{ odd},j_{1}\textrm{ even},r(i+j+j_{1})=-1$ & $IB(i)=IS1N(i)$\tabularnewline
\hline 
$IS1N(i)>0,j\textrm{ odd},j_{1}\textrm{ odd},r(i+j+j_{1})=1$ & $IB(i)=IS1N(i)$\tabularnewline
\hline 
$IS2N(i)>0,j\textrm{ even},j_{1}\textrm{ even},r(i+j+j_{1})=-1$ & $IB(i)=IS2N(i)$\tabularnewline
\hline 
$IS2N(i)>0,j\textrm{ even},j_{1}\textrm{ odd},r(i+j+j_{1})=1$ & $IB(i)=IS2N(i)$\tabularnewline
\hline 
$IS2N(i)>0,j\textrm{ odd},j_{1}\textrm{ even},r(i+j+j_{1})=1$ & $IB(i)=IS2N(i)$\tabularnewline
\hline 
$IS2N(i)>0,j\textrm{ odd},j_{1}\textrm{ odd},r(i+j+j_{1})=-1$ & $IB(i)=IS2N(i)$\tabularnewline
\hline 
\end{tabular}\\
\\
\\
\begin{tabular}{|c|c|}
\hline 
If & Then also\tabularnewline
\hline 
\hline 
$IS1N(i)>0,ID(i+j+j_{1})>0,$ & $IB(i)=IS1N(i)$\tabularnewline
$j\textrm{ even},j_{1}\textrm{ even},r(i+j+j_{1})=-1$ & \tabularnewline
\hline 
$IS1N(i)>0,ID(i+j+j_{1})>0,$ & $IB(i)=IS1N(i)$\tabularnewline
$j\textrm{ even},j_{1}\textrm{ odd},r(i+j+j_{1})=1$ & \tabularnewline
\hline 
$IS1N(i)>0,ID(i+j+j_{1})>0,$ & $IB(i)=IS1N(i)$\tabularnewline
$j\textrm{ odd},j_{1}\textrm{ even},r(i+j+j_{1})=1$ & \tabularnewline
\hline 
$IS1N(i)>0,ID(i+j+j_{1})>0,$ & $IB(i)=IS1N(i)$\tabularnewline
$j\textrm{ odd},j_{1}\textrm{ odd},r(i+j+j_{1})=-1$ & \tabularnewline
\hline 
$IS2N(i)>0,ID(i+j+j_{1})>0,$ & $IB(i)=IS1N(i)$\tabularnewline
$j\textrm{ even},j_{1}\textrm{ even},r(i+j+j_{1})=1$ & \tabularnewline
\hline 
$IS2N(i)>0,ID(i+j+j_{1})>0,$ & $IB(i)=IS1N(i)$\tabularnewline
$j\textrm{ even},j_{1}\textrm{ odd},r(i+j+j_{1})=-1$ & \tabularnewline
\hline 
$IS2N(i)>0,ID(i+j+j_{1})>0,$ & $IB(i)=IS1N(i)$\tabularnewline
$j\textrm{ odd},j_{1}\textrm{ even},r(i+j+j_{1})=-1$ & \tabularnewline
\hline 
$IS2N(i)>0,ID(i+j+j_{1})>0,$ & $IB(i)=IS1N(i)$\tabularnewline
$j\textrm{ odd},j_{1}\textrm{ odd},r(i+j+j_{1})=1$ & \tabularnewline
\hline 
\end{tabular} 
\end{enumerate}
Although the \emph{normal} and additional rules cover most cases,
there are still a few rare cases where the correct answer is not found.
A more exact method will be described later, when considering distortions.
In practice one can use the normal rules only when the error-rate
is low, use the normal and additional rules when the error-rate is
higher, but refer to a more exact (and more lengthy) calculation described
later if the noise is very high and especially in presence of a strong
and correlated interference.

\subsection{Binary error-rate without precoding}

With all the previous rules, the binary error-rate without precoding
(or, with precoding, the error-rate on the precoded sequence) can
be computed by the following algorithm:
\begin{itemize}
\item Consider all the possible duobinary transmitted sequences starting
and ending with a symbol $1$ or $-1$ of length $L=3,\,4,\,5,\,\ldots$
\item Consider, for each transmitted sequence, all the possible recognized
sequences and compute the duobinary error probability, by writing
the system of inequalities corresponding to the duobinary tests and
solving this system by numerical integrations.
\item Count the number of simple and double errors on the survivors and
then the number of simple errors transformed into double errors, by
applying the normal and the additional rules.
\end{itemize}
The total probability of binary errors is then the sum of the computed
duobinary error probabilities multiplied by the number of final double
errors and weighted by the probabilities of the considered sequences.
Fortunately the number of sequences introducing binary errors is limited,
which means that the majority of duobinary errors are eliminated during
the convergence of the survivors.

\begin{figure}[tbph]
\begin{centering}
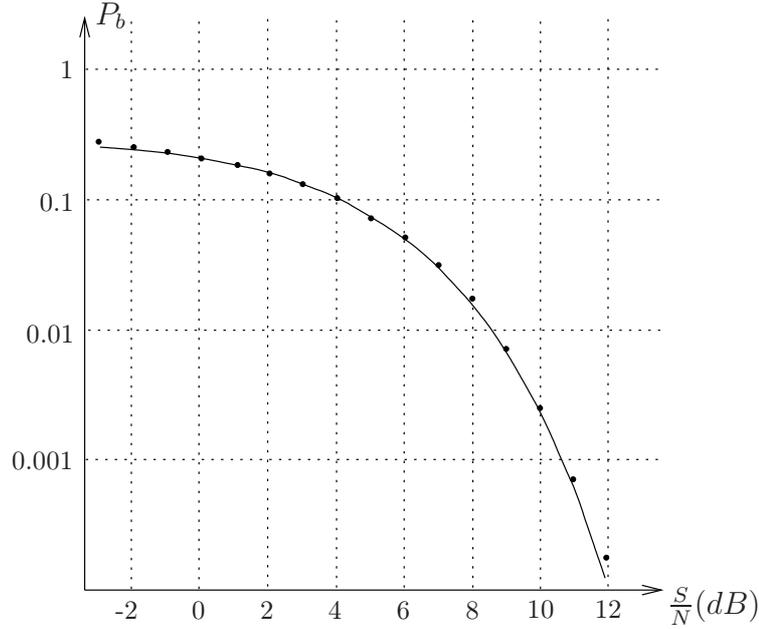
\par\end{centering}
\caption{Binary error-rate as a function of $S/N$ without precoding: theoretical
curve and points of simulation.\label{figure9}}
\end{figure}

The above algorithm was followed to obtain the binary error-rate.
It was found that it is not necessary to consider sequences of length
greater than 6, their contribution to the final result being negligible.
The theoretical curve of the binary error-rate as a function of $S/N$
is given in Figure~\ref{figure9} together with the results of simulation.

\subsection{Binary error-rate with precoding}

In order to obtain the binary error-rate with precoding, we have to
note that a single binary error on the precoded sequence produces
two binary errors on the input sequence, but that $k$ consecutive
errors on the precoded sequence only produce two errors too on the
input sequence. The calculation procedure is then similar to the procedure
described in the previous Section with two changes.

First we assume that each final double error on the survivors gives
two binary errors on the input sequence, but this leads to an overestimated
number of binary errors. Let $q$ be the probability of a double error
on the survivors. Consecutive double errors can occur $2,\,3,\,\ldots$
times with probabilities $q^{2},\,q^{3},\,\ldots$ It is easy to show
that we must subtract, from $q$, the number of errors in excess which
is
\begin{equation}
2q^{2}+4q^{3}+6q^{4}+\ldots\comma
\end{equation}
so that the binary error probability becomes
\begin{equation}
q_{p}=2\left[q-2q^{2}(1+2q+3q^{2}+4q^{3}+...)\right]\comma
\end{equation}
which is equal to 
\begin{equation}
q_{p}=2\left[q-\frac{2q^{2}}{(1-q)^{2}}\right]\point
\end{equation}

\begin{figure}[tbph]
\begin{centering}
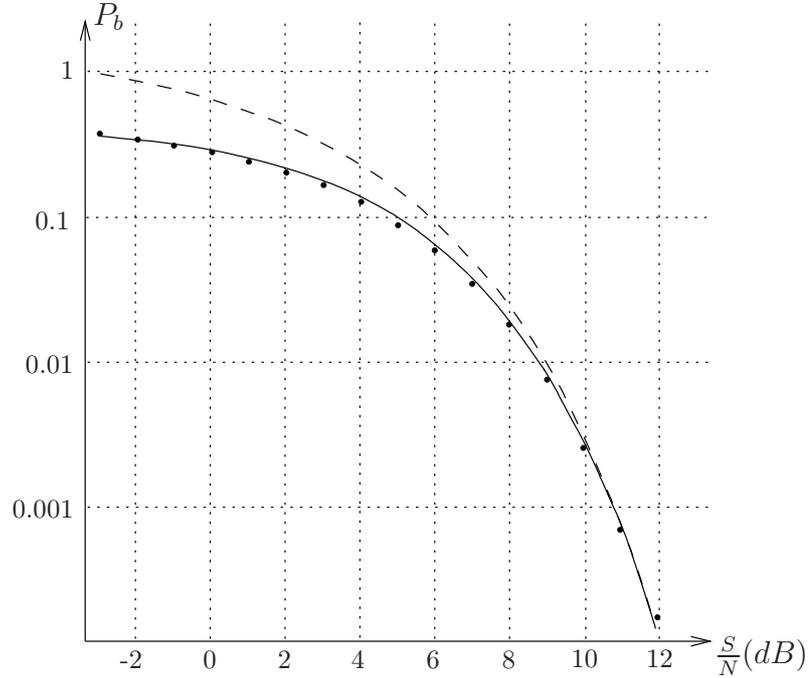
\par\end{centering}
\caption{Comparison of methods giving the binary error-rate as a function of
$S/N$ with precoding: theoretical curve (full line), simulations
(points), and theoretical upper bound as established by Allard and
detailed in Section~\ref{subsec:Upper-bound-of} (dashed line).\label{figure10}}
\end{figure}

The mean binary error-rate is finally computed by a weighted summation
of the $q_{p}$ terms corresponding to the different duobinary sequences.
Theoretical values of the binary error-rate were computed as just
described and compared with results obtained by simulation. All these
values have been drawn in Figure~\ref{figure10}.

This figure also shows the results of the theoretical upper bound
of the BER developed by Alard in~\cite{UER-Doc-GTR5-101} which is
summarized in the next Section.

\subsection{Upper bound of the binary error-rate\label{subsec:Upper-bound-of}}

The EBU document~\cite{UER-Doc-GTR5-101} reproduces a study made
by Alard at the CCETT (France) where an upper bound of the BER is
derived for the duobinary code. As this document is not generally
available, a summary of the theoretical calculations of the bound
is given hereafter.

Let us denote by $s_{k}$ the path correctly followed in the trellis
(\ie the sequence of the binary input states), by $s'_{k}$ the path
recognized by the decoder, by $x_{k}$ and $x'_{k}$ the duobinary
sequences transmitted and recognized, and by $y_{k}$ the received
sequence impaired by the noise of variance $\sigma^{2}$.

It is assumed that the path recognized is in general identical to
the correct path but that it can diverge from the later immediately
after time $t$ and reconverge immediately before time $t+r$ (otherwise
the code would be subject to ``catastrophic error propagation''
which has no practical interest). Such an event of divergence followed
by reconvergence of the paths is denoted by $\epsilon$ and is called
as a \emph{closed error event}, by opposition to an \emph{open error
event} where there is no reconvergence of the paths until infinity
and therefore catastrophic propagation.

The squared Euclidean distance between the transmitted and recognized
duobinary sequences is defined by
\begin{equation}
D^{2}(\varepsilon)=\sum_{k=t+1}^{k=t+r-1}(x_{k}-x'_{k})^{2}\comma\label{eq:SquareEuclideanD}
\end{equation}
which is shown to be an even integer larger or equal to $2$.

It is then assumed that the recognized sequence $x'_{k}$ is, among
all the possible sequences $x_{i}$, the closest to the received sequence
$y_{k}$. In other words, $x'_{k}$ is at a lower distance of $y_{k}$
than any other possible sequence and in particular the correct sequence
$x_{k}$.

Since the decision distance is $D(\epsilon)/2$, the probability of
an closed error event $\epsilon$ is bounded by 
\begin{equation}
P(\varepsilon)\leq Q\left(\frac{D(\varepsilon)}{2\sigma}\right)\point\label{eq:Pepsilon}
\end{equation}

If considering only confusion between adjacent duobinary symbols,
thus neglecting the less probable confusion between $-1$ and $+1$,
the squared Euclidean distance (\ref{eq:SquareEuclideanD}) reaches
its minimum value equal to $2$.

It remains to count the number of duobinary sequences that are at
the squared Euclidean distance $2$ from the transmitted sequence
$x_{k}$. If the path $s'_{k}$ diverges from the correct path after
instant $t$ and reconverges just before instant $t+r$, the duobinary
symbols $x'_{t+1}$ and $x'_{t+r-1}$ are different from $x_{t+1}$
and $x_{t+r-1}$, which is sufficient to reach the minimum squared
distance $2$. This is only possible if the state sequence $s_{k}$
takes alternate values $0$ and $1$ (or $-1$ and $+1$), the adversary
sequence $s'_{k}$ taking the opposite values.

Therefore there is only one path diverging from the correct path from
$t+1$ to $t+r-1$ with the corresponding duobinary sequences $x_{k}$
and $x'_{k}$ being at the minimum squared Euclidean distance $2$.
The probability of existence of this path is $(1/2)^{r-2}$. It follows
that, for the whole set $\mathcal{E}$ of $\epsilon$ events, the
probability $P(\mathcal{E})$ is, when making $D(\epsilon)$ in~(\ref{eq:Pepsilon})
equal to its minimum value which is $\sqrt{2}$, 

\begin{equation}
P(\mathcal{E})\leq\sum_{r=2}^{\infty}\left(\frac{1}{2}\right)^{r-2}Q\left(\frac{\sqrt{2}}{2\sigma}\right)=2\,Q\left(\frac{1}{\sigma\sqrt{2}}\right)\point\label{eq:PETotal}
\end{equation}

Without precoding an $\epsilon$ event produces $r-1$ binary errors,
while with precoding it produces $2$ binary errors.

Finally the upper bound of binary error-rate is:
\begin{itemize}
\item without precoding,
\begin{equation}
P_{b}\leq\sum_{r=2}^{\infty}\,\frac{r-1}{2^{r-2}}\,Q\left(\frac{1}{\sigma\sqrt{2}}\right)=4\,Q\left(\frac{1}{\sigma\sqrt{2}}\right)\comma\label{eq:PbwithoutPrecoding}
\end{equation}
\item and with precoding,
\begin{equation}
P_{b}\leq\sum_{r=2}^{\infty}\,\frac{2}{2^{r-2}}\,Q\left(\frac{1}{\sigma\sqrt{2}}\right)=4\,Q\left(\frac{1}{\sigma\sqrt{2}}\right)\point\label{eq:PbwithPrecoding}
\end{equation}
\end{itemize}
Both (identical) expressions, developed by Allard, give a good approximation
of the BER for high values of $S/N$ but overestimate the error probability
for low $S/N$; this is clearly shown in the calculations by bounding
the probability of an $\epsilon$ event in equation (\ref{eq:Pepsilon})
and by neglecting the errors produced by confusion between duobinary
symbols $-1$ and $+1$. 

As both equations (\ref{eq:PbwithoutPrecoding}) and (\ref{eq:PbwithPrecoding})
provide the same result, it has been thought that the BER is the same
with and without precoding. Although the error-rate becomes closer
and closer to the value without precoding for high $S/N$, our results
show that the error-rate is always higher without precoding. As seen
in Figure~\ref{figure10}, equations (\ref{eq:PbwithoutPrecoding})
and (\ref{eq:PbwithPrecoding}) clearly give a good approximation
for the low noise channels ($S/N$ high) but they become insufficient
for channels impaired by strong noise.

The paper by Altekar \etal~\cite{Altekar1999Error} presents a different
approach based on the partial response codes defined by a polynomial
of the form
\begin{equation}
h(D)=(1-D)^{m}(1+D)^{n}\point\label{eq:AltekarPolynomial}
\end{equation}
In this paper, an input error sequence is defined as 
\begin{equation}
\varepsilon_{s}(D)=s(D)-s'(D)\comma
\end{equation}
where $s$ and $s'$ are respectively the correct states and the states
recognized by the \viterbi decoder in binary numbers $0$ and $1$.
An output error sequence is then 
\begin{equation}
\varepsilon_{y}(D)=h(D)\,\varepsilon_{s}(D)\point
\end{equation}

The performance of the system is largely dictated by input error sequences
that result in an output with small squared Euclidean distance 
\begin{equation}
\left\Vert \varepsilon_{y}(D)\right\Vert ^{2}=\sum_{k}\,\varepsilon_{y,\,k}^{2}\point
\end{equation}
The paper describes two algorithms, on the basis of a so-called \emph{error
state diagram}, used to derive all the input error sequences corresponding
to a closed error event (where the paths $s$ and $s'$ diverge after
time $t$ and reconverge at time $t+r$) and to an open error event
(where the paths do not reconverge). The input error sequences corresponding
to a given squared Euclidean distance are finally listed for several
codes, \ie for different values of $m$ and $n$ in (\ref{eq:AltekarPolynomial}).

For $m=0$ and $n=1$, the code considered in the paper is the duobinary
code which is the subject of our study. The first input error sequence
listed for a closed error event and the minimum squared Euclidean
distance equal to $2$ is 
\begin{equation}
(0),\,1,\,(-1,\,1,\,-1,\,1,\,\ldots)\,0\point
\end{equation}

Recalling that this is the difference $s(D)-s'(D)$, we can obtain
$s$ and $s'$ separately according to the following table 
\begin{center}
\begin{tabular}{|c|c|c|}
\hline 
$s-s'$ & $s$ & $s'$\tabularnewline
\hline 
\hline 
$0$ & $0$ & $0$\tabularnewline
\hline 
$-1$ & $0$ & $-1$\tabularnewline
\hline 
$1$ & $1$ & $0$\tabularnewline
\hline 
$0$ & $1$ & $1$\tabularnewline
\hline 
\end{tabular}
\par\end{center}

Therefore the input error sequence given corresponds to the following
states: 
\begin{eqnarray}
s & = & 1,\,(0,\,1,\,0,\,1,\,\ldots)\\
s' & = & 0,\,(1,\,0,\,1,\,0,\,\ldots)\point
\end{eqnarray}

This is similar to the conclusion of Allard since during the interval
of paths divergence, the correct state sequence $s$ takes alternate
values $0$ and $1$ while the adversary sequence $s'$ takes the
opposite values. Moreover if the input error sequence is duobinary
coded in order to obtain the output error sequence, we have
\begin{equation}
1\,(0,\,0,\,0,\,0,\,\ldots)\,1\comma
\end{equation}
which shows that the total squared Euclidean distance is $2$.

Altekar \etal\emph{~}\cite{Altekar1999Error} give another input
error event with a squared Euclidean distance $2$ which leads to
the same conclusion so that, when considering only the minimum squared
Euclidean distance, the bounds of the binary error-rate given by (\ref{eq:PbwithoutPrecoding})
and (\ref{eq:PbwithPrecoding}) are confirmed. The same paper also
gives other input error sequences with squared Euclidean distances
of $6$ and $10$; these sequences could avoid neglecting confusion
between duobinary symbols $-1$ and $+1$ which will make the previous
bounds tighter. 

\subsection{Error statistics\label{subsec:Error-statistics}}

The duobinary errors are practically independent. On the other hand,
the binary errors are certainly not independent since we have seen
that a single duobinary error produces $n+1$ consecutive survivors
errors, $n$ being the number of symbols $0$ correctly recognized
by the decoder before the duobinary error. One must therefore expect
that a \viterbi decoder, with and without precoding, will give errors
in bursts. With precoding, one single binary error as well as a burst
of consecutive errors on the precoded sequence produces two final
binary errors. It is then expected that with precoding the bursts
are somewhat less frequent than without precoding.

To quantify this behavior we have computed by simulation the probabilities
of 2, 3 and 4 consecutive binary errors. If $p$ is the probability
of binary errors and if these errors were independent, the probability
of $k$ consecutive errors would be $p^{k}$, while the actual probabilities
of $k$ consecutive errors are higher. Let $p_{k}$ denote the probability
of $k$ consecutive errors. Table~\ref{table4} gives, in rounded
figures, the ratio $p_{k}/p^{k}$ for $k=2,\,3,\,4$. 
\begin{table}[tbph]
\begin{centering}
\begin{tabular}{|c|c|c|c|c|c|c|}
\hline 
$S/N$ & \multicolumn{3}{c|}{Without precoding} & \multicolumn{3}{c|}{With precoding}\tabularnewline
$(\decibel)$ & $k=2$ & $k=3$ & $k=4$ & $k=2$ & $k=3$ & $k=4$\tabularnewline
\hline 
\hline 
$0$ & $1.5$ & $2.5$ & $4$ & $1.5$ & $1.5$ & $2.5$\tabularnewline
\hline 
$6$ & $10$ & $65$ & $5$ & $5$ & $5$ & $25$\tabularnewline
\hline 
$9$ & $65$ & $4\times10^{3}$ & $525$ & $33$ & $30$ & $730$\tabularnewline
\hline 
\end{tabular}
\par\end{centering}
\caption{Ratio $p_{k}/p^{k}$ of the actual probability of $k$ consecutive
binary errors to the same probability for independent errors, with
$k=2,\,3,\,4$.\label{table4}}
\end{table}

It is seen that binary errors occur in longer and longer bursts when
$S/N$ increases. Short bursts are however less frequent with precoding.

\section{Interferences\label{sec:Interferences}}

This section is devoted to the study of the effect of several types
of interferences and distortions on the performance of a \viterbi
decoder. 

In the following, we will define as an \emph{additive} \emph{interference}
(or simply as an \emph{interference}), anything which is linearly
added to the signal, except noise. In general, interference is the
result of a linear distortion in the transmission chain, but it can
also come from \emph{crosstalk} between different channels.

If $d(t)$ is the duobinary signal (\ie the duobinary coded version
of the data sequence) and if there is an interference $y(t)$, the
input level without noise of the \viterbi decoder is 
\begin{equation}
x(t)=d(t)+y(t)\comma\label{equation58}
\end{equation}
while, in presence of noise, it is
\begin{equation}
x(t)=d(t)+y(t)+n(t)\point
\end{equation}

\subsection{Correlated or uncorrelated interference}

As shown in Section~\ref{subsec:Error-statistics}, in presence of
white noise, of which the successive samples are independent, the
duobinary errors are also quasi-independent while the binary errors
occur in bursts.

If the samples of the interference are uncorrelated, error-rates can
be computed as done for the case of pure noise, simply by replacing
noise by the sum of noise plus interference. The calculation method
is again to define, by application of the duobinary tests, a system
of simultaneous inequalities where $n_{i}$ becomes $n_{i}+y_{i}$,
and to derive the corresponding probability by numerical integrations,
noting that the values of the interference will appear in the limits
of integrations.

In most cases however, the interference samples are correlated and
the \emph{duobinary} errors will also occur in bursts. For example,
let us consider the case where the wanted duobinary signal $d_{1}(t)$
is interfered by another sequence $d_{2}(t)$ which is: 
\begin{itemize}
\item a duobinary sequence, independent of $d_{1}$, or
\item a ternary sequence taking the same levels $-1,\,0,\,1$ as a duobinary
sequence with the same probabilities but without correlation.
\end{itemize}
In absence of noise, the probabilities of 2, 3 and 4 consecutive duobinary
errors, obtained by simulation, are given in Table~\ref{table5}.
\begin{table}[tbph]
\begin{centering}
\begin{tabular}{|c|c|c|c|}
\hline 
Cases & 2 errors & 3 errors & 4 errors\tabularnewline
\hline 
\hline 
$d_{2}$ = correlated duobinary sequence  & $0.194$ & $0.0757$ & $0.0366$\tabularnewline
\hline 
$d_{2}$ = uncorrelated ternary sequence  & $0.123$ & $0.0409$ & $0.0146$\tabularnewline
\hline 
\end{tabular}
\par\end{centering}
\caption{Probabilities of 2, 3 and 4 consecutive duobinary errors for an interference
$y=d_{2}$.\label{table5}}
\end{table}

It is seen that the probabilities of consecutive duobinary errors
are significantly increased by the correlation of the interference,
with two consequences:
\begin{itemize}
\item the \viterbi decoding is no longer optimal because the metrics in
the trellis are no more strictly additive. However \viterbi decoding
still gives a significant improvement with reference to threshold
decoding.
\item the theoretical error-rates can no longer be computed by the method
of numerical integration of a system of inequalities. However this
method is still valid as a first approximation and gives good results
when the interference is low and is not strongly correlated.
\end{itemize}

\subsection{General theoretical method in absence of noise\label{subsec:General-theoretical-method-no-noise-18}}

A general algorithm which should give the exact theoretical error-rates,
involves the following steps:
\begin{enumerate}
\item Consider the duobinary initial transmitted sequences of length $3,\,4,\,5,\ldots$,
starting and ending by a symbol $1$ or $-1$. \\
For example
\[
L=3\,\,\,\,\,\,\,\,\,\,1\,0\,-1\,,\,-1\,0\,1\,,\,1\,1\,1\,,\,-1\,-1\,-1
\]
\[
\textrm{with probability of the intermediate symbol }1/8
\]
\[
L=4\,\,\,\,\,\,\,\,\,\,1\,0\,0\,1\,,\,-1\,0\,0\,-1\,,\,1\,0\,-1\,-1\,,\,-1\,0\,1\,1
\]
\[
\textrm{with probability of the intermediate symbols }1/16
\]
etc. The probabilities $1/8,$ $1/16$,~\ldots{} quoted for the intermediate
symbols come from the fact that these duobinary symbols are the coded
form of binary sequences with length $4$ for $L=3$, of length $5$
for $L=4$, and so on. Since the input bits $1$ and $-1$ have the
same probability, the probability of a binary sequence of length $4$
is $1/8$, of length $5$ is $1/16$,~\ldots{} Observe that for any
length $L$, there are only four duobinary sequences starting and
ending by $1$ or $-1$, so that the total probability of intermediate
symbols in all given duobinary sequences is $1$. Therefore all possible
cases are considered. 
\item In the case of noise (or uncorrelated interference), interference
is applied to all symbols of these sequences but it was assumed that
the first $1$ (or $-1$) and the last $1$ (or $-1$) are correctly
recognized. Here we cannot avoid the case when there is a duobinary
error on these symbols, as well as on the foregoing and the following
symbols. We will then prolong the initial sequence to the left and
to the right by known symbols $1$ (or $-1$) in order to have an
interfered sequence of length $L_{t}$. We will again prolong these
sequences to the left and to the right by $L_{c}$ symbols $1$ (or
$-1$) $L_{c}$ being the decoding constraint length of the decoder.
The total length of the transmitted sequence then becomes $L_{t}+2L_{c}$
as illustrated hereafter for the first initial sequences of length
3.\\
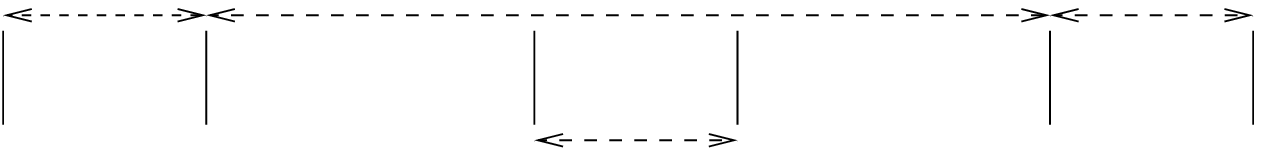
\item Apply successively all possible values of the interference to the
$L_{t}$ symbols of the interfered sequence, but leave the first and
the last $L_{c}$ symbols free of interference.
\item Apply the duobinary tests given by (\ref{equation27}) to (\ref{equation30})
to the whole sequence to obtain the successive duobinary symbols $r_{i}$
recognized by decoder. The presence of the first $L_{c}$ non interfered
symbols allows a correct initialization of this process, while the
last $L_{c}$ symbols allows complete termination of the decoding
which is made with a delay of $L_{c}$ symbols or bits.
\item Store and count all the cases where there is a duobinary error on
the last but one symbol of the initial sequence. If there is no error
on this symbol, iterate directly to the next value of the interference.
\item From the recognized duobinary symbols, reconstruct the survivors by
using the rules of \emph{copying} or \emph{crossing} the survivors
given in Section~\ref{sec:Duobinary-theoretical-framework}.
\item Go to the next step of \viterbi decoding and follow the evolution
of the survivors.
\item Steps 3 to 7 above should be made with the actual values of the interference
and also with no interference at all; in this second case there will
be no duobinary nor binary errors, so that it is possible to construct
two matrices (for the survivors $S_{1}$ and $S_{2}$) with interference
and two similar matrices for $S_{1}$ and $S_{2}$ without interference
and thus error free.
\item Compare the survivors matrices with and without errors and detect,
then count the binary errors after $L_{c}$ steps of the decoding
(\ie after convergence of the survivors).
\end{enumerate}
The binary error-rate is finally the sum of all binary errors detected
weighted by the probabilities of the values considered for the interference
and also weighted by the probabilities of the intermediate symbols
in the initial duobinary transmitted sequence.

In principle, this method should provide the exact answer if
\begin{itemize}
\item the length $L$ of the initial sequences is $3,\,4,\,\ldots,\,\infty$.
\item the length $L_{t}$ of the interfered sequence is infinite.
\item the constraint length (assuring complete convergence of the survivors)
is infinite. In practice some truncation of this triple infinity is
obviously needed and there should be some compromise between the precision
required and the available memory as well as the time of computation.
We adopted the figures $L=3,\,4,\,\ldots,\,8$, $L_{t}=11$, $L_{c}=10$,
which give a good \emph{second} \emph{approximation} of the error-rates
(our first approximation being the method of numerical integration).
\end{itemize}

\subsection{General theoretical method in presence of interference and noise\label{subsec:General-theoretical-method-19}}

The computation algorithm is the same as in the preceding Section,
except that, at Step 3, it is necessary to combine each possible value
of the interference with each possible value of the noise.

Suppose that with $L_{t}=11$, we have $2^{11}=2048$ values for the
interfering sequences. If we sample the Gaussian distribution of noise
into 100 values and store their probabilities, we will end up with
$2048^{100}$ possible values of the total perturbation interference
+ noise. Such a number of cases is unmanageable. The only practical
way is to replace all values or noise by a sufficiently high number
of random sequences of Gaussian noise. We decided to generate 1000
noise sequences, which, as will be shown, gives the correct order
of magnitude, but the final theoretical curves have then to be slightly
smoothed.

For the same reason, when the number of values of the interference
is too large, we will limit the number of simulations to 1000 random
sequences of interference.

We will now consider the cases where the interference is the result
of:
\begin{enumerate}
\item a phase error $\varphi$ in a coherent demodulator of CQPRS modulation, 
\item a synchronization error $\tau$ in the receiver, and 
\item an echo of relative amplitude $\beta$, of phase $\psi$ and of delay
$\tau$ introduced in the radio-frequency path in CQPRS or AM/SSB
modulations.
\end{enumerate}

\subsection{Interference resulting from a phase error $\varphi$ in a CQPRS demodulation}

\subsubsection{Phase error without noise}

If $d_{1}(t)$ and $d_{2}(t)$ are the two duobinary sequences modulating
the carriers in quadrature and if there is a phase error in the first
demodulator, the input level of the \viterbi decoder is
\begin{equation}
x(t)=d_{1}(t)\cos\varphi-d_{2}(t)\sin\varphi\point
\end{equation}
According to the definition (\ref{equation58}), the interference
is given by
\begin{equation}
y(t)=-d_{1}(t)(1-\cos\varphi)-d_{2}(t)\sin\varphi\point
\end{equation}
There is therefore a mutual interference between the two independent
duobinary sequences $d_{1}$ and $d_{2}$ vanishing for $\varphi=0$.

For a given length $L$, there are $2^{L}$ possible values for the
\emph{interfering} sequence $d_{2}$, given by duobinary coding of
$2^{L}$ input bits $\pm1$. Since the input bits are assumed to be
independent and have the same probability, all the duobinary sequence
$d_{2}$ have the same probability of $1/(2^{L})$. For the case of
$L_{t}=11$, considered in our second approximation there are $2^{11}=2048$
interfering sequences $d_{2}$.

\begin{figure}[tbph]
\begin{centering}
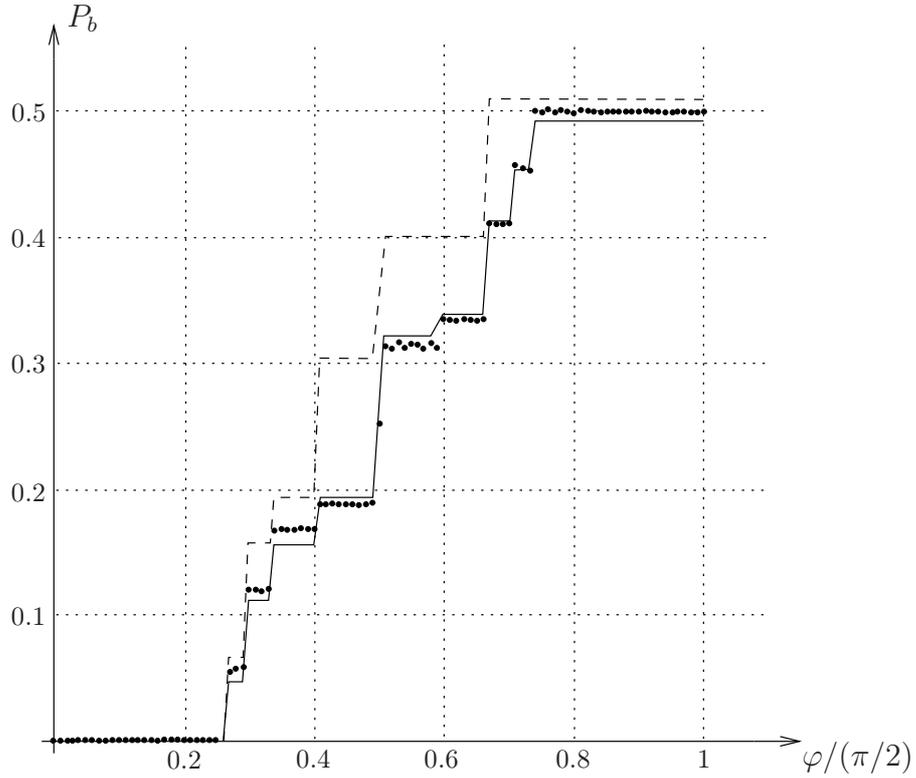
\par\end{centering}
\caption{Comparison of methods giving the binary error-rate with a phase error
$\varphi$ in CQPRS without noise: first approximation (dashed line),
second approximation (full line), and simulations (points).\label{figure11}}
\end{figure}

As explained above, a first approximation of the error-rates is obtained
by numerical integration with the transmitted sequences $d_{1}$ used
in Section~\ref{subsec:Duobinary-error-rate-10} and all the $2^{L}$
values of $d_{2}$. A second theoretical approximation is obtained
by the general method described in Section~\ref{subsec:General-theoretical-method-no-noise-18}.
Results of both approximations are given in Figure~\ref{figure11}
for the binary error-rate, together with the results of simulation,
for $\varphi$ varying from $0$ to $\pi/2$. The case of a phase
error $\varphi$ in CQPRS demodulation, without noise, with duobinary
coding, and with \viterbi decoding (unmatched filtering), was chosen
for Figure~\ref{figure11}. 

It can be seen that the first approximation is reasonably accurate
for the duobinary error-rate and the survivors error-rate, even for
large $\varphi$. For the binary error-rate we need the second approximation
which takes account of the correlation of $d_{1}$ and $d_{2}$ by
considering longer interfered sequences.

All the error-rates exhibit sudden variations for specific values
of $\varphi$ and then remain constant up to another \emph{threshold}
of variation. In particular the binary error-rate is strictly zero
for $\varphi\leq0.27\,\pi/2$. These thresholds of variation correspond
to limits in the duobinary tests. For example, test (\ref{equation27})
can here be written as
\begin{equation}
(d_{1,\,i}+d_{1,\,i-k})\cos\varphi-(d_{2,\,i}+d_{2,\,i-k})\sin\varphi>1\,\,\,\,\,\textrm{or}\,\,\,\,\,<0\point\label{equation62}
\end{equation}
The limit $1$ of this test is reached if
\begin{equation}
\varphi\approx0.27\,\frac{\pi}{2}\,\,\,\,\,\textrm{or}\,\,\,\,\,\cos\varphi-\sin\varphi=\frac{1}{2}\comma
\end{equation}
and with $d_{1,\,i-k}=d_{1,\,i}=d_{2,\,i-k}=d_{2,\,i}=1$.

Similarly the other thresholds of $\varphi$ are found to be
\begin{equation}
\varphi\approx0.30\,\frac{\pi}{2}\,\,\,\,\,\textrm{or}\,\,\,\,\,\cos\varphi-2\sin\varphi=0
\end{equation}
\begin{equation}
\varphi\approx0.34\,\frac{\pi}{2}\,\,\,\,\,\textrm{or}\,\,\,\,\,\sin\varphi=\frac{1}{2}
\end{equation}
\begin{equation}
\varphi\approx0.41\,\frac{\pi}{2}\,\,\,\,\,\textrm{or}\,\,\,\,\,\cos\varphi+2\sin\varphi=1
\end{equation}
\begin{equation}
\varphi\approx0.5\,\frac{\pi}{2}\,\,\,\,\,\textrm{or}\,\,\,\,\,\cos\varphi-\sin\varphi=0
\end{equation}
\begin{equation}
\varphi\approx0.6\,\frac{\pi}{2}\,\,\,\,\,\textrm{or}\,\,\,\,\,-\cos\varphi+\sin\varphi=1
\end{equation}
\begin{equation}
\varphi\approx0.71\,\frac{\pi}{2}\,\,\,\,\,\textrm{or}\,\,\,\,\,2\cos\varphi-\sin\varphi=0
\end{equation}
\begin{equation}
\phi\approx0.74\,\frac{\pi}{2}\,\,\,\,\,\textrm{or}\,\,\,\,\,\cos\varphi-\sin\varphi=\frac{1}{2}\point
\end{equation}

\subsubsection{Phase error with noise}

\begin{figure}[htbp]
\begin{centering}
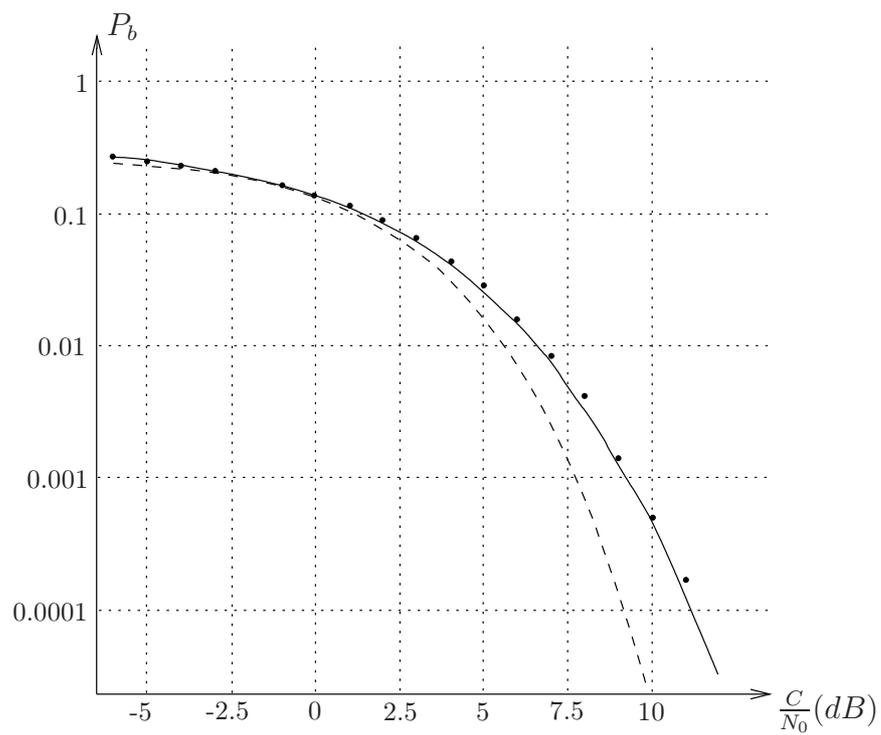
\par\end{centering}
\caption{Comparison of methods giving the binary error-rate with a phase error
$\varphi=0.1\,\pi/2$ in CQPRS with noise as a function of $C/N_{0}$:
first approximation (dashed line), second approximation (full line),
and simulations (points). \label{figure12}}
\end{figure}

\begin{figure}[htbp]
\begin{centering}
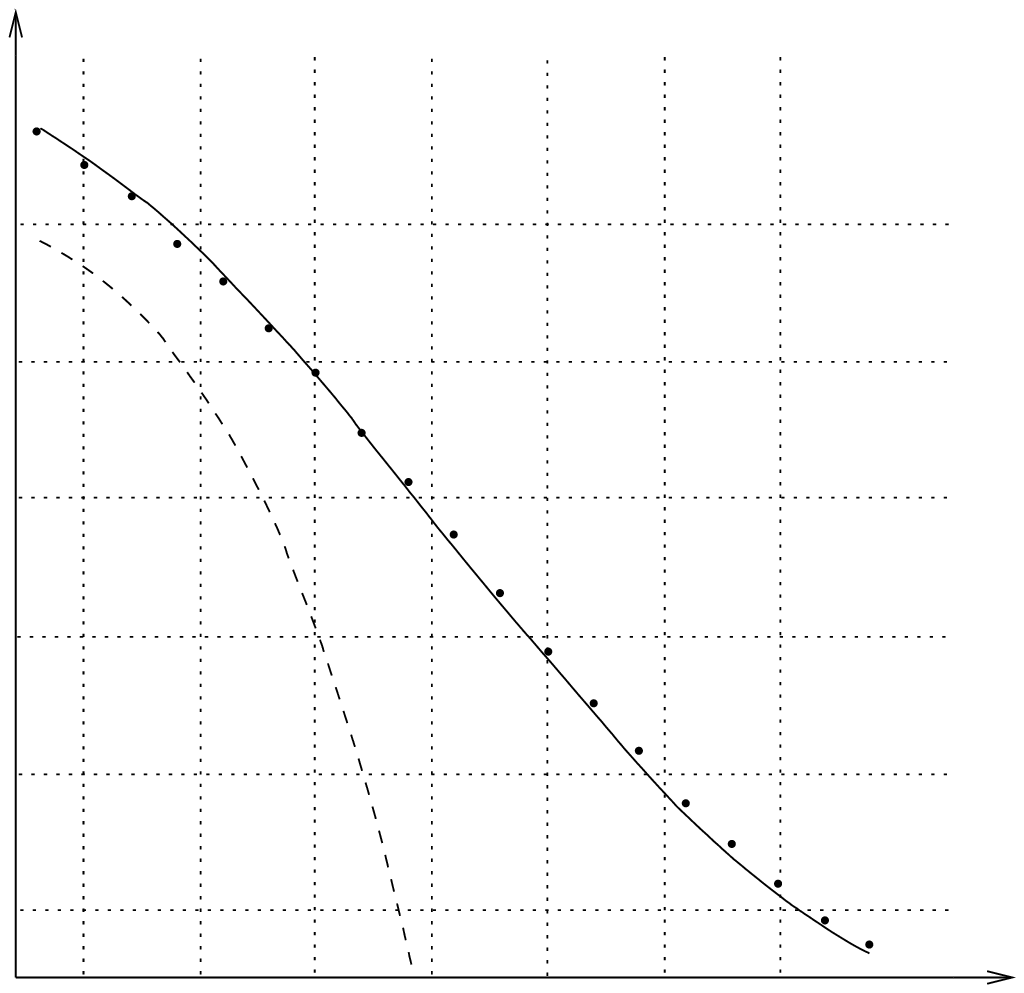
\par\end{centering}
\caption{Comparison of methods giving the binary error-rate with a phase error
$\varphi=0.28\,\pi/2$ in CQPRS with noise as a function of $C/N_{0}$:
first approximation (dashed line), second approximation (full line),
and simulations (points).\label{figure13}}
\end{figure}

Two examples of theoretical binary error-rates, compared with simulation
are given in Figures~\ref{figure12} and~\ref{figure13}, as functions
of the ratio $C/N_{0}$ (carrier power to RF noise spectral density
in CQPRS). Figure~\ref{figure12} assumes a phase error of $0.1\,\pi/2$
(\ie 9 degrees) while Figure~\ref{figure13} assumes a phase error
of $0.28\,\pi/2$, which is already above the first threshold of variation.
Again, theoretical values obtained by the second approximation described
in Section~\ref{subsec:General-theoretical-method-19} are very close
to the simulation results.

\subsection{Interference resulting from a synchronization error $\tau$}

\subsubsection{Synchronization error without noise}

If, due to an improper synchronization recovery, the duobinary signal
is sampled at times $kT+\tau$ instead of $kT$, the signal level
at decoder input is 
\begin{equation}
x(kT+\tau)=d(kT+\tau)\comma
\end{equation}
and there is an interference
\begin{equation}
y(kT+\tau)=-d(kT)+d(kT+\tau)\point
\end{equation}
The calculation starts with the evaluation of the level of $x(kT+\tau)$.
This problem has been solved by constructing a file of all possible
levels to obtain the second approximation of the eye diagram (see
Section~\ref{subsec:Duobinary-eye-diagram-ex4}). Depending on the
transmitted sequence (including the duobinary symbols before and after
$d(kT)$), the level $x$ is simply red in the file at the appropriate
address.

The error-rates can then be computed by the general algorithm of Section~\ref{subsec:General-theoretical-method-no-noise-18},
but some care must be taken here with regards to the sign of $\tau$.
On the one hand, examination of the eye diagram shows a complete symmetry
for a positive or negative value of $\tau$. On the other hand, in
theoretical calculation, one must remember that the duobinary errors
on the \emph{critical} symbol (last but one symbol of the initial
transmitted sequence of length $L$) depend on duobinary symbols transmitted
\emph{before} the critical symbol, as expressed by the terms $d_{i-k}$
in the duobinary tests. Therefore the theoretical calculation should
be made with $\tau<0$ but the result can be applied, by symmetry,
for $\tau>0$.

In a more detailed discussion of this problem, a distinction should
be made between the duobinary errors on a symbol $0$ and on a symbol
$1$ (or $-1$). It can also be shown by a discussion of the duobinary
tests similar to that explained by equation (\ref{equation62}) for
a phase error, that the binary error-rate, without noise, is strictly
zero for any value of $|\tau|<T/2$, \ie half of the bit period.
One can then already predict that \viterbi decoding will ensure a
good robustness against synchronization errors.

\begin{figure}[tbph]
\begin{centering}
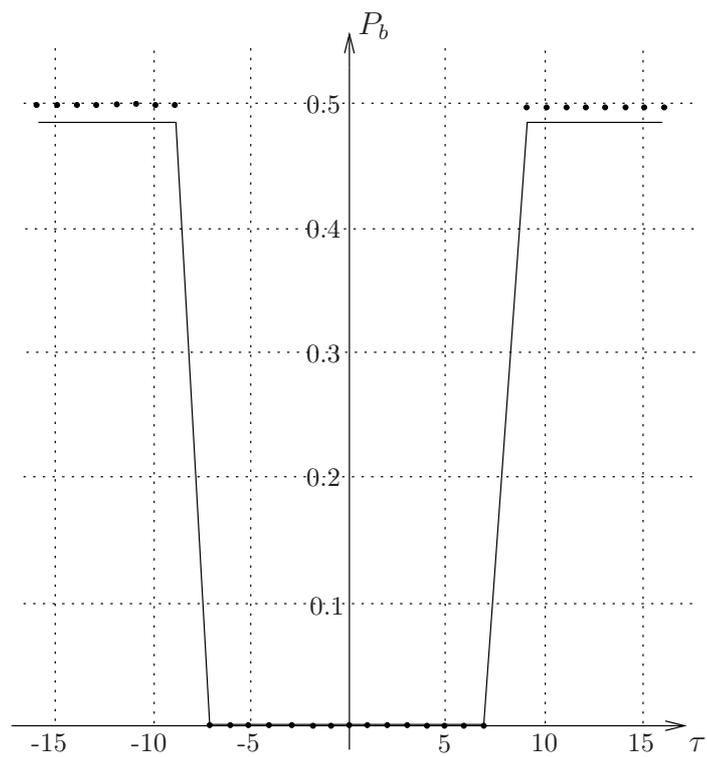
\par\end{centering}
\caption{Binary error-rate with a synchronization error $\tau$ in baseband
and without noise: theoretical curve and points of simulation. Variation
of $\tau$: from $-15T/16$ to $15T/16$. \label{figure14}}
\end{figure}

The theoretical binary error-rate compared with simulation is given
in Figure~\ref{figure14}, where a bit period $T$ is divided in
16 parts. It is seen that this error-rate is zero until $\tau=8T/16$
and then jumps to $1/2$.

\subsubsection{Synchronization error with noise}

\begin{figure}[tbph]
\begin{centering}
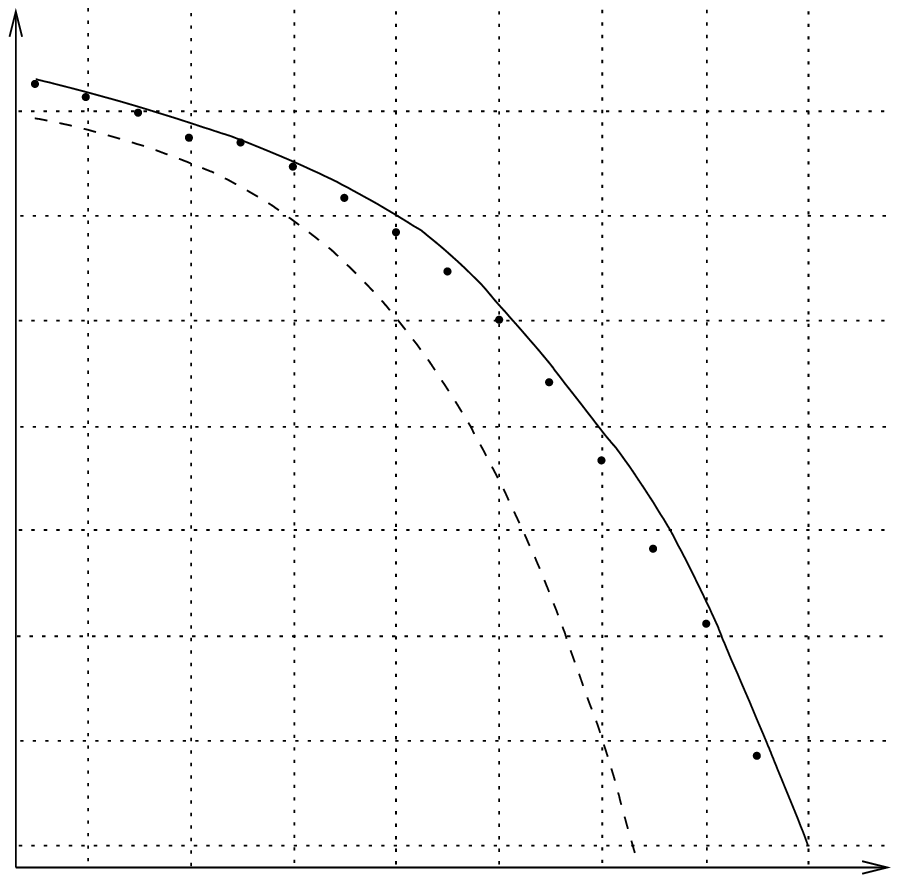
\par\end{centering}
\caption{Comparison of methods giving the binary error-rate with a synchronization
error $\tau=4T/16$ in baseband and with noise as a function of $S/N$:
theory without synchronization error (dashed line), second theoretical
approximation (full line), and simulations (points). \label{figure15}}
\end{figure}

For a small value of $\tau$ (up to $2T/16$), the first approximation
by numerical integration gives the correct order of magnitude of the
binary error-rate. For larger $\tau$, it is necessary to use the
general algorithm of Section~\ref{subsec:General-theoretical-method-19}.
One example of theoretical binary error-rates is given in Figure~\ref{figure15}
for $\tau=4T/16$. It can also be shown that for $\tau=2T/16$ (\ie
$T/8$) the binary error-rate is not very much increased with reference
to $\tau=0$.

\subsection{Interference resulting from an echo in CQPRS}

\subsubsection{Echo without noise}

If $d_{1}(t)$ and $d_{2}(t)$ are again the two duobinary sequences
modulating the two carriers in CQPRS and if, on the radio-frequency
path there is an echo of relative amplitude $\beta$, of delay $\tau$
and of phase $\psi$, the input level at the \viterbi decoder is
\begin{equation}
x(t)=d_{1}(t)+\beta\,[d_{1}(t-\tau)\cos\psi-d_{2}(t-\tau)\sin\psi]\comma
\end{equation}
$d_{1}(t)$ being considered as the wanted signal. The interference
is simply the echo signal
\begin{equation}
y(t)=\beta\,[d_{1}(t-\tau)\cos\psi-d_{2}(t-\tau)\sin\psi]\point\label{equation74}
\end{equation}
The first duobinary test (\ref{equation27}) is written as
\begin{equation}
d_{1,\,i}+d_{1,\,i-k}+y_{i}+y_{i-k}\,\,>0\,\,\,\,\,\textrm{OR}\,\,\,<1\comma
\end{equation}
with similar expressions for the three other tests. If $\beta$ is
zero or very small, there should be no binary errors at all, but as
$\beta$ becomes larger and larger there could be binary errors due
to interference, even without noise. In order to find the minimum
value $m$ of $\beta$ which gives binary errors, let us see when
the test (\ref{equation62}) reaches the limit $1$. If $\tau$ is
a multiple of $T$, this condition is fulfilled with $d_{1,\,i}(0)=1$,
$d_{1,\,i-k}(0)=1$, $d_{1,\,i}(-kT)=-1$, $d_{1,\,i-k}(-kT)=-1$
and becomes 
\begin{equation}
2-2\beta=1\comma
\end{equation}
so that 
\begin{equation}
\beta_{m}=\frac{1}{2}\point
\end{equation}
If now $\tau$ is still a multiple of $T$ but $\psi=\pi/4$ (or an
odd multiple of $\pi/4$), we have
\begin{equation}
2-2\beta\sqrt{2}=1
\end{equation}
\begin{equation}
\beta_{m}=\frac{1}{2\sqrt{2}}=0.3535\point
\end{equation}

Finally, if we maximize the interference with respect to $\tau$ and
$\psi$, we must have $\tau=(2k+1)T$ and $\psi=(2k+1)\pi/4$ and
then, by taking the values of $d_{1}(t-\tau)$ and $d_{2}(t-\tau)$
in the file of levels (described in Section~\ref{subsec:Duobinary-eye-diagram-ex4}),
we obtain
\begin{equation}
\beta_{m}=\frac{1}{2.414\sqrt{2}}=0.293\approx-10.7\,\decibel\point
\end{equation}
Therefore, in practice, there will be no binary errors at all (in
absence of noise) if the echo is lower than $-10.7\,\decibel$. This
is confirmed by simulations.

\subsubsection{Echo with noise}

In the presence of noise, the error-rates are function of four parameters,
namely the $C/N_{0}$ ratio and the echo parameters $\beta$, $\tau$,
$\psi$. Let us analyze this multidimensional surface.

\begin{figure}[tbph]
\begin{centering}
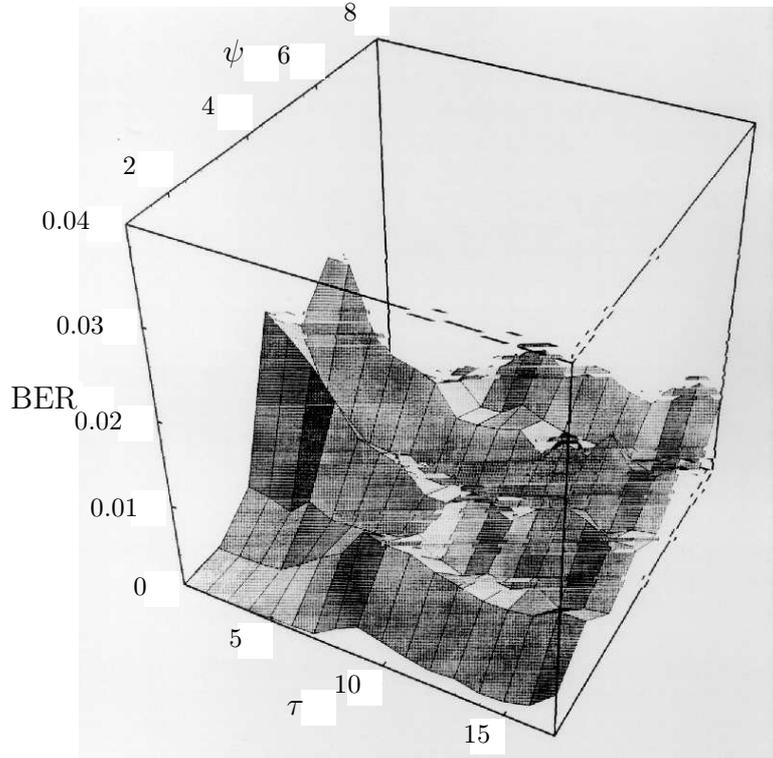
\par\end{centering}
\caption{Binary error-rate with an echo in CQPRS modulation as a function of
echo delay $\tau$ and echo phase $\psi$: echo delay $\tau=0$ to
$4T$ \ie $\tau=(x-1)T/4$, echo phase $\psi=0$ to $7\pi/4$ \ie
$\psi=(y-1)\pi/4$, and echo amplitude $\beta=0.31$, $C/N_{0}=12\,\decibel$.
\label{figure16}}
\end{figure}

Figure~\ref{figure16} is a cross-section of this surface corresponding
to $\beta=0.31$ and $C/N_{0}=12\,\decibel$ obtained by simulation.
It is seen that the binary error-rate is maximum for very short echo
delays ($\tau\leq2T$) and for phases between $\pi$ and $3\pi/2$.
This could be expected because the first term of interference (\ref{equation74})
is then strongly correlated with the direct signal. For longer delays
($\tau\geq3T$), the cross-sections become practically flat and the
bit-error rate depends only slightly on $\tau$ and $\psi$. However
if $C/N_{0}$ is high, or in other words if the effect of interference
is bigger then the effect of noise, the binary error-rate tends to
become a periodical function of $\tau$, with period $T$ with small
variations, and also a periodical function of $\psi$, with period
$\pi/4$ and somewhat bigger variations, the critical phases being
odd multiples of $\pi/4$. This is clearly shown in Figure~\ref{figure16}.

\begin{figure}[tbph]
\begin{centering}
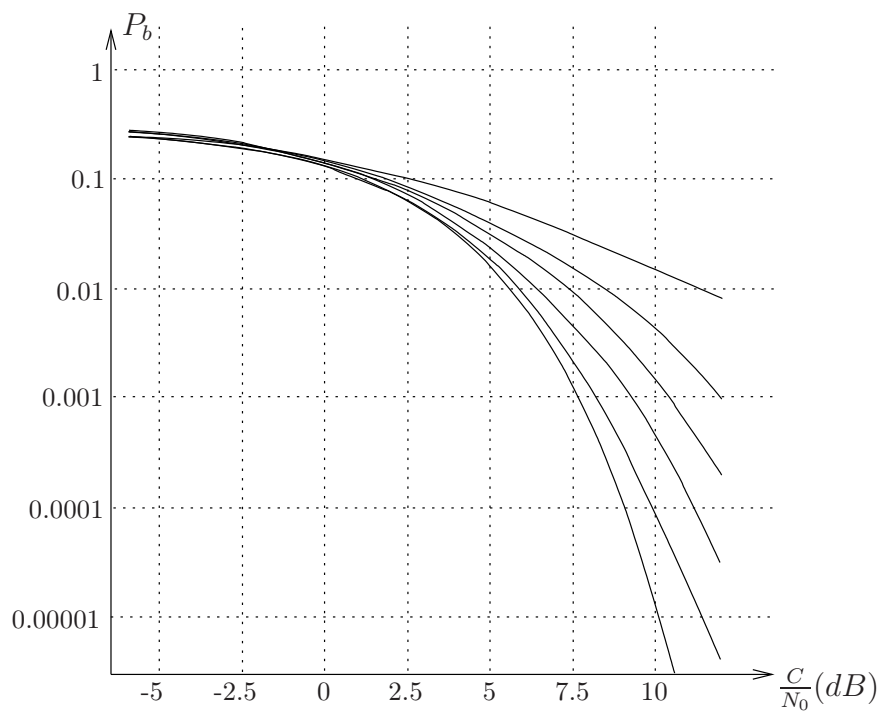
\par\end{centering}
\caption{Average binary error-rate with an echo in CQPRS modulation as a function
of $C/N_{0}$ and echo amplitude $\beta$: $\beta=0,\,0.10,\,0.15,\,0.20,\,0.25,\,0.30$
(in order of increasing error-rate), echo phase $\psi=\pi/4$, and
echo delay $\tau$ = long and multiple of $T$.\label{figure17}}
\end{figure}

A first approximation of the theoretical binary error-rate can again
be obtained by numerical integrations. The second approximation, with
the general method of Section~\ref{subsec:General-theoretical-method-19},
gives better results, but as the error-rate does not depend very much
on $\tau$ and $\psi$ (for $\tau$ large), it is possible to define
an \emph{average} binary error-rate depending only on $C/N_{0}$ and
$\beta$. Choosing for $\psi$ the value $\pi/4$ makes this average
slightly pessimistic. The results of the theoretical calculation (second
approximation) is given in Figure~\ref{figure17} for $\beta=0$
(no echo), $0.10$, $0.15$, $0.20$, $0.25$ and $0.31$. This single
Figure thus immediately gives the order of magnitude of the error-rate
for a given echo amplitude, as a function of $C/N_{0}$. 

\subsection{Interference resulting from an echo in AM/SSB}

\subsubsection{Echo without noise}

As continuous amplitude modulation with vestigial sideband (AM/VSB)
is used for distribution of broadcasting signals over cable networks,
we have considered AM/SSB for the transmission of a digital duobinary
signal, as an approximation to AM/VSB. 

If $p$ and $q$ are the amplitude of the lower and upper sidebands
($p$ and $q$ are functions of frequency), the input level at the
decoder in presence of an echo is
\begin{eqnarray}
x(t) & = & (p+q)\,[d(t)+\beta\,d(t-\tau)\,\cos\psi]\nonumber \\
 &  & +(p+q)\,[\beta\,d_{q}(t-\tau)\,\sin\psi]\comma
\end{eqnarray}
where $d_{q}(t)$ is the signal in quadrature with $d(t)$.

The extreme cases are double sideband (AM/DSB) where $p=q=1/2$ at
all frequencies and single sideband (AM/SSB) where $p=1$ and $q=0$
(or $p=0$ and $q=1$). We will examine the case of AM/SSB which is
close to AM/VSB with a simpler expression since
\begin{equation}
x(t)=d(t)+\beta\,d(t-\tau)\,\cos\psi+\beta\,d_{q}(t-\tau)\,\sin\psi\point
\end{equation}
We have explained in Section~\ref{subsec:Duobinary-eye-diagram-ex4}
that all possible levels of the signal $d_{q}(t)$ in quadrature with
a duobinary signal $d(t)$ were also stored in a computer file. The
procedure (reading the level at the appropriate address) can then
be used for $d_{q}(t)$ as well as for $d(t)$.

The interference is given by
\begin{equation}
y(t)=\beta\,d(t-\tau)\,\cos\psi+\beta\,d_{q}(t-\tau)\,\sin\psi\point
\end{equation}
The first duobinary test (\ref{equation27}) takes the form
\begin{equation}
\begin{array}{c}
d_{i}(kT)+d_{i-k}(kT)+\beta\,\cos\psi\,[d_{i}(kT-\tau)+d_{i-k}(kT-\tau)]\\
+\beta\,\sin\psi\,[d_{i,q}(kT-\tau)+d_{i-k,q}(kT-\tau)]\,\,\,<0\,\,\,\,\,\textrm{OR}\,\,\,>1\point
\end{array}\label{equation84}
\end{equation}
The minimum value $\beta_{m}$ of $\beta$ which gives binary errors
in absence of noise is found by the same method as used for a CQPRS
echo, by expressing the condition where the test (\ref{equation84})
reaches the limit $1$.

If $\psi=0$ or $\psi=\pi$, and if $\tau$ is a multiple of $T$,
we obtain
\begin{equation}
\beta_{m}=\frac{1}{2}\point
\end{equation}

When $\psi$ is still equal to $0$ or $\pi$, but if we maximize
the first term $d(kT-\tau)$ of the interference with respect to the
delay $\tau$, the file of the levels of $d$ indicates that $\tau=(2k+1)T/2$
with a level of $1.207$, so that
\begin{equation}
\beta_{m}=0.4142\point\label{equation86}
\end{equation}

If now $\psi=\pi/2$ or $3\pi/2$, the interference is only due to
$d_{q}$ which has a maximum amplitude of $1.6677$ for a delay $\tau$
multiple of $T$, so that
\begin{equation}
\beta_{m}=0.2981\point\label{equation87}
\end{equation}

For other cases, such as $\psi=(2k+1)\pi/4$, the values of $\beta_{m}$
are between (\ref{equation86}) and (\ref{equation87}). Therefore,
in rounded figures, there will never be binary errors in absence of
noise if
\begin{equation}
\beta<0.2981\approx-10.5\,\decibel\point
\end{equation}

\subsubsection{Echo with noise}

In the case of an echo with noise, the binary error-rate is a multidimensional
surface function of four parameters, namely the $S/N$ ratio and the
echo characteristics $\beta$, $\tau$, $\psi$.

\begin{figure}[tbph]
\begin{centering}
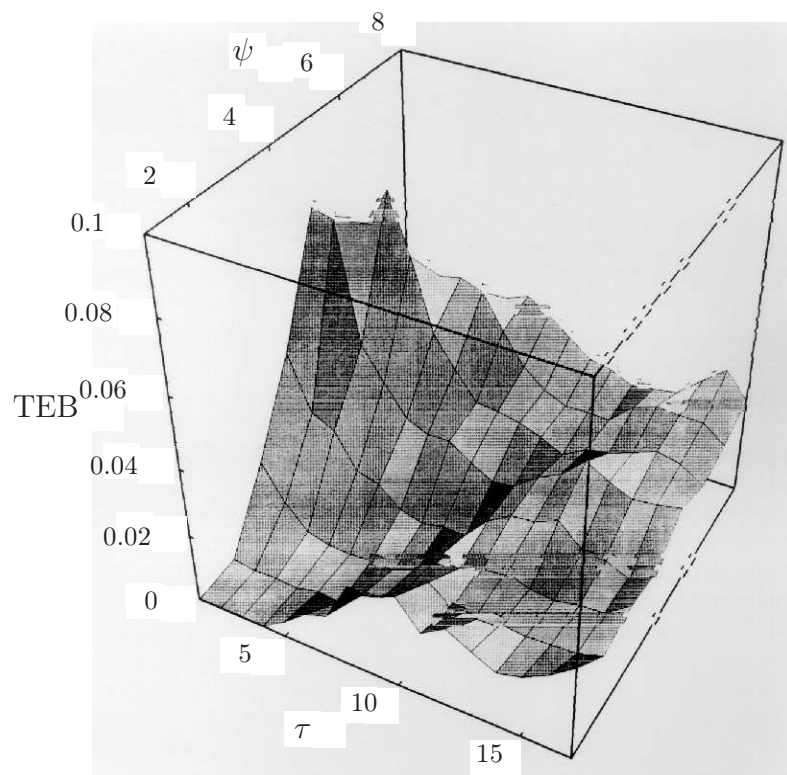
\par\end{centering}
\caption{Binary error-rate with an echo in AM/SSB modulation as a function
of echo delay $\tau$ and echo phase $\psi$: echo delay $\tau=0$
to $4T$ \ie $\tau=(x-1)T/4$, echo phase $\psi=0$ to $7\pi/4$
\ie $\psi=(y-1)\pi/4$, and echo amplitude $\beta=0.31$, $S/N=12\,\decibel$.\label{figure18}}
\end{figure}

Figure~\ref{figure18}, obtained by simulation, is a cross-section
of this surface corresponding to $\beta=0.31$ and $S/N=12\,\decibel$.
It is seen that the error-rate is very high for short delays ($\tau\leq6T$),
where not only $d(t-\tau)$ but also the quadrature component $d_{q}(t-\tau)$
are heavily intercorrelated with the direct signal. The critical phases
are between $\pi$ and $3\pi/2$. For longer delays, the cut becomes
approximately flat and it is again possible to define an \emph{average}
binary error-rate, preferably with the most critical phase which is
$3\pi/2$. 

As was done for the echo in CQPRS, a first theoretical approximation
of the binary error-rate can be obtained by numerical integrations,
especially for low interference. A second theoretical approximation
is obtained by the general method of Section~\ref{subsec:General-theoretical-method-19}
and is valid even for strong echoes.

\begin{figure}[tbph]
\begin{centering}
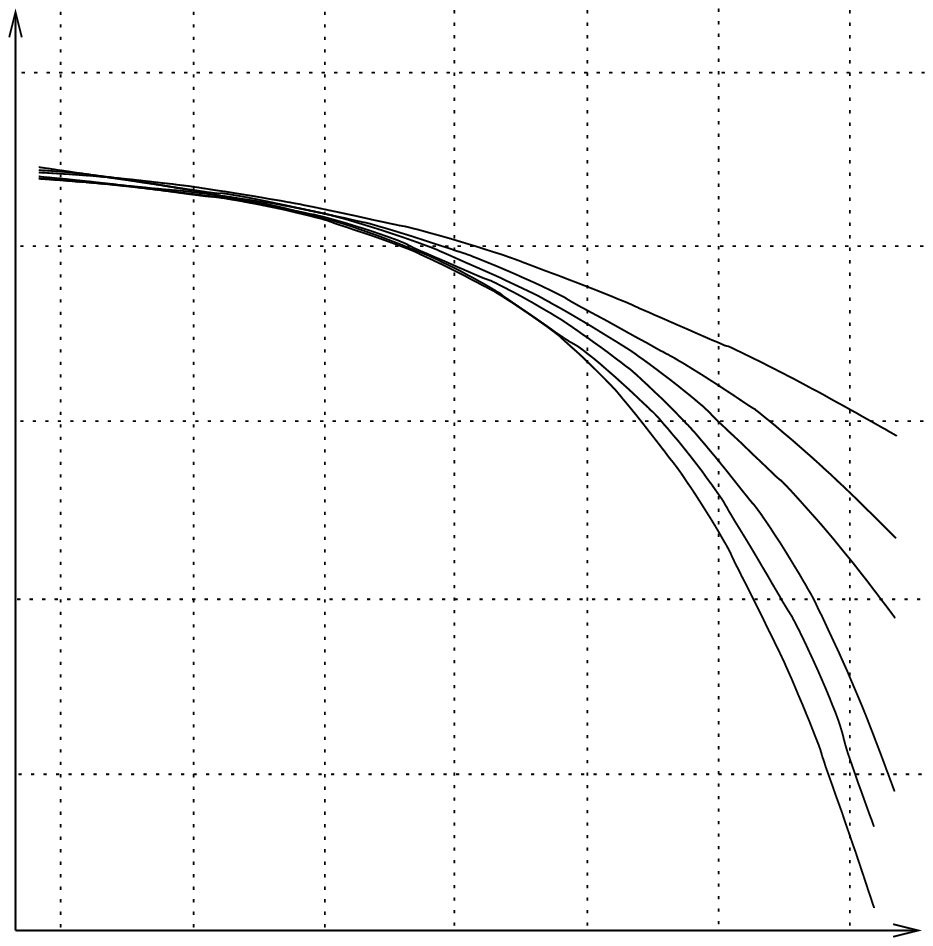
\par\end{centering}
\caption{Average binary error-rate with an echo in AM/SSB modulation as a function
of $S/N$ and echo amplitude $\beta$: $\beta=0,\,0.10,\,0.15,\,0.20,\,0.25,\,0.30$
(in order of increasing error-rate), echo phase $\psi=3\pi/2$, echo
delay $\tau$ = long and multiple of $T$.\label{figure19}}
\end{figure}

Finally, Figure~\ref{figure19} gives the \emph{average} binary error-rates
for $\beta=0$, $0.1$, $0.15$, $0.20$, $0.25$, and $0.31$. 

As far as echoes with short delays are concerned (consider for example
a system with a bit-rate of $10~\text{Mbit/s}$), a delay of $T$
corresponds to a path difference of $30$ meters which is not unlikely
on a cable network. Care must therefore be taken to prevent echoes
with short delays in cable distribution.

\subsection{Example of a non-linear distortion}

\subsubsection{General}

The previous methodology can be extended to the case of a non-linear
distortion of known characteristics. For the discussion, we have selected
the example of a traveling-wave tube (TWT\nomenclature{TWT}{Traveling-Wave Tube})
having typical AM/AM and AM/PM characteristics. The TWT is considered
as the high power amplifier for QPRS modulation with coherent demodulation,
but for comparison we have also studied the case of the off-set modulation
COQPRS.

\subsubsection{Non-linearity model}

In front of the non-linear device, the modulated signal can be written
as
\begin{equation}
x(t)=\rho(t)\,e^{j\theta(t)}\comma
\end{equation}
while after passing through the non-linear device, it becomes
\begin{equation}
z(t)=\rho_{d}(t)\,e^{j[\theta(t)+\varphi(t)]}\point
\end{equation}
 When adopting the so-called ``Saleh memoryless model'', the AM/AM
characteristic is
\begin{equation}
\rho_{d}(t)=A_{s}^{2}\,\frac{\rho(t)}{\rho^{2}(t)+A_{s}^{2}}\comma\label{equation91}
\end{equation}
and the AM/PM characteristic is
\begin{equation}
\varphi(t)=\frac{\pi}{3}\,\frac{\rho^{2}(t)}{\rho^{2}(t)+A_{s}^{2}}\comma\label{equation92}
\end{equation}
where $A_{s}$ is the TWT saturation level.

\begin{figure}[tbph]
\begin{centering}
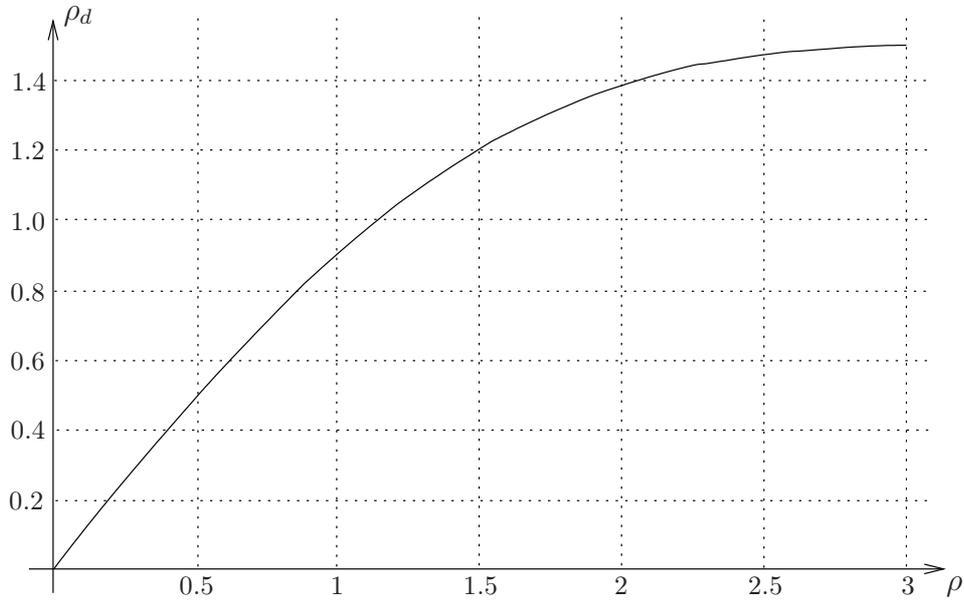
\par\end{centering}
\caption{AM/AM characteristics of the traveling-wave tube: saturation level
$A_{s}=3$.\label{figure20}}
\end{figure}

\begin{figure}[tbph]
\begin{centering}
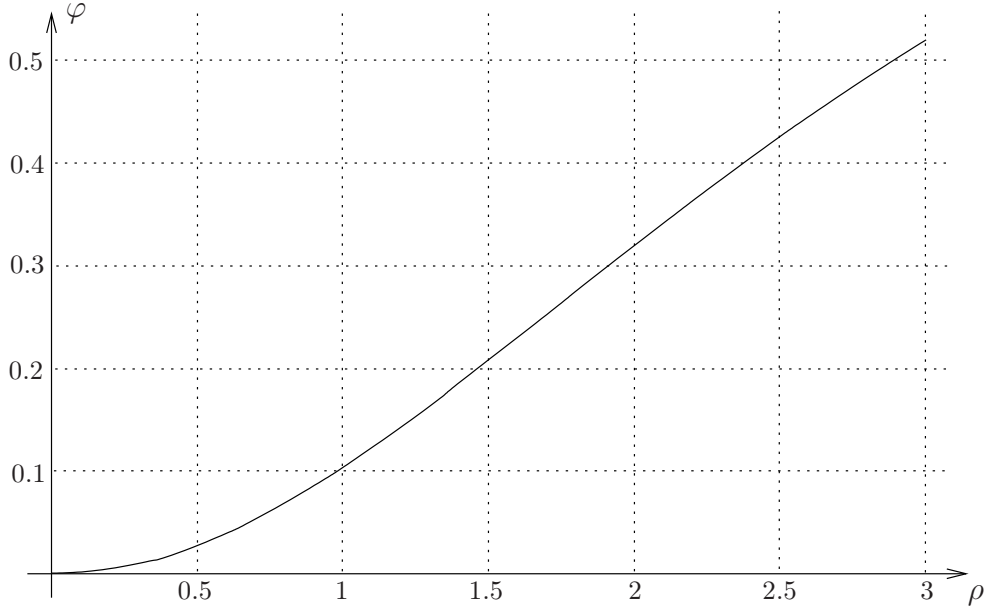
\par\end{centering}
\caption{AM/PM characteristics of the traveling-wave tube: saturation level
$A_{s}=3$.\label{figure21}}
\end{figure}

In the following, we select the value $A_{s}=3$ which corresponds
to a significant but not exaggerated distortion for a carrier amplitude
$A_{c}=1$. The AM/AM and AM/PM characteristics with $A_{s}=3$ are
reproduced in Figures~\ref{figure20} and~\ref{figure21}.

\subsubsection{Conventional definition of the back-off}

The saturation level $A_{s}$ is measured at the \emph{input} of the
TWT. According to the AM/AM characteristic of equation (\ref{equation91})
and Figure~\ref{figure20}, the maximum \emph{output} amplitude is
$A_{s}/2$ and the maximum \emph{output} power is $A_{s}^{2}/4$.
The usual definition of the backoff $B_{\textrm{off}}$ is the ratio
of the maximum output power to the actual output power
\begin{equation}
B_{\textrm{off}}=\frac{P_{\textrm{out max}}}{P_{\textrm{out}}}=\frac{A_{s}^{2}}{4\,P_{\textrm{out}}}\comma
\end{equation}
$P_{\textrm{out }}$ being the mean value of $z^{2}(t)$.\\

In order to facilitate the calculations, we will adopt a slightly
different definition. Since we consider only 4-state modulations,
we note that the maximum output power on one channel, for example
the channel corresponding to the real part of $x(t)$ and $z(t)$,
is $A_{s}^{2}/8$. Then if $C$ is the output power that would be
obtained on one channel in the \emph{absence of distortion} (\ie
with a \emph{linear} TWT of power gain equal to $1$ and a carrier
amplitude $A_{c}=1$), we define the backoff as the ratio
\begin{equation}
B_{\textrm{off}}=\frac{P_{\textrm{out max }(\textrm{one channel})}}{C}=\frac{A_{s}^{2}}{8C}\point
\end{equation}
We will compute the error-ratio without distortion (linear operation)
and with distortion (non-linear operation) as a function of the same
parameter $C/N_{0}$ and assess the penalty due to the distortion
in terms of $C/N_{0}$.

In our tests, we have used the following numerical values of the parameters
for CQPRS and unmatched filtering 
\begin{equation}
C=\frac{A_{c}^{2}}{4}=0.25\comma
\end{equation}
\begin{equation}
B_{\textrm{off}}=6.53\,\decibel\comma
\end{equation}
and a noise variance $\sigma^{2}$ computed by 
\begin{equation}
\frac{1}{\sigma^{2}}=2\,\frac{C}{N_{0}}\,T_{s}\point
\end{equation}

\subsubsection{Note on band-pass filtering}

In principle a band-pass filter must be inserted at the input and
another band-pass filter at the output of the TWT. However, for CQPRS,
there are no spectral components outside the RF band
\begin{equation}
B=f_{b}\comma
\end{equation}
so that the first band-pass filter is useless.

In all cases, the non-linearity introduces a certain amount of spectrum
spreading. It can however be shown that, with the parameters used
for our tests, the components outside the band $B$ are very small
(about 1\% relative to the centre of the band). We will therefore
neglect also the second band-pass filtering.

\subsubsection{Method of calculation}

For each particular modulation, the first step is to write the series
of values taken by the real and imaginary parts $x_{r}(t)$ and $x_{i}(t)$
of the modulated signal at the sampling instants $kT$ which are multiples
of the symbol period $T$ if $t=0$ corresponds to the middle of one
symbol. These values are then weighted with their respective probabilities.

The second step is to compute the values of the envelope $\rho(kT)$
and the phase $\theta(kT)$ at the sampling instants, before distortion.
The next step is to apply equations (\ref{equation91}) and (\ref{equation92})
to obtain the envelope $\rho_{d}(kT)$ and the phase $\theta(kT)+\varphi(kT)$,
after distortion.

The final step is to compute the real part $z_{r}(kT)$ of the distorted
signal by
\begin{equation}
z_{r}(kT)=\rho_{d}(kT)\,\cos\left[\theta(kT)+\varphi(kT)\right]\point
\end{equation}

The difference
\begin{equation}
y_{r}(kT)=z_{r}(kT)-x_{r}(kT)\label{equation100}
\end{equation}
can then be considered as an additive interference and we can derive
the binary error-rate on the \emph{real} part of the channel by using
the methods given above. Since there is a symmetry between the \emph{real}
part of the channel $z_{r}(t)$ and the \emph{imaginary} part of the
channel $z_{i}(t)$, the error-rate is the same on both channels.
If needed, $z_{i}(kT)$ can be computed by
\begin{equation}
z_{i}(kT)=\rho_{d}(kT)\,\sin\left[\theta(kT)+\varphi(kT)\right]\comma
\end{equation}
with an interference
\begin{equation}
y_{i}(kT)=z_{i}(kT)-x_{i}(kT)\point
\end{equation}

\subsection{CQPRS modulation with \viterbi decoding, no precoding and unmatched
filtering}

We have already seen that the \viterbi decoding, although no longer
optimal in presence of interference, still gives a good improvement
over threshold decoding. This is also the case in presence of a non-linear
distortion.

Two general methods were presented to compute the binary error-rate.
Both methods involve: 
\begin{enumerate}
\item the calculations of the series of transmitted duobinary sequences
of length $L=3,\,4,\,5,\,\ldots$, starting and finishing by a duobinary
symbol $1$ or $-1$, 
\item the application of the duobinary tests in order to detect duobinary
errors on the last but one symbol and,
\item computation of the consequential binary errors.
\end{enumerate}
In the first method, this is done by numerical integrations to obtain
the probability of a system of simultaneous inequalities which contains
the successive noise samples.

This method does not take account of the correlation of the interference
and thus gives only a first approximation of the final error-rate,
if, as it is the case in CQPRS, the interference is coming from a
independent duobinary sequence. Note that the values of interference
(or distortion) here appear in the limits of the integrations.

We will only present here the results with the second method which
gives a better approximation. In this second method, the transmitted
sequence is prolonged by known symbols and, after the duobinary tests,
the survivors are reconstructed so that the binary errors are detected
by following their evolution. Account is taken here of the correlation
of interference.

We again consider all duobinary transmitted sequences of length $L=3,\,\ldots,\,8$
starting and finishing by $1$ or $-1$, but these sequences are prolonged
on each side by known symbols $1$ or $-1$ up to length $L_{t}=11$.
For each transmitted sequence, we apply all possible interfering sequences
(\ie $2^{11}=2048$) to the $11$ symbols $L_{t}$ as well as a sufficient
number or random noise sequences. In addition, on each side of the
$L_{t}=11$ symbols, the sequence is prolonged by $L_{c}$ known symbols
$1$ or $-1$, but without adding noise or interference. This will
allow the correct initialization and termination of the series of
duobinary tests. All the duobinary symbols are then correctly recognized
and the survivors can be reconstructed by the \emph{recopying} rule
at each step of the decoding.

The evolution of the survivors is then followed, once with noise and
distortion and once without noise or distortion. The comparison of
the two series of survivors gives the correct number of binary errors
in the sense that we count the number of binary errors due to duobinary
errors on the last but one symbol of the initial transmitted sequence.
In each case, the interference $y(i,j)$ is computed according to
a relation similar to (\ref{equation100}) in which $z_{r}$ is written
$z_{r}(i,j)$ where $i$ is the iteration on the wanted sequence and
$j$ the iteration on all possible duobinary interfering sequences.

As explained previously, this method is exact, provided that the lengths
$L$, $L_{t}$ and $L_{c}$ are infinite and that an infinite number
of noise sequences is used. In practice truncation is necessary to
limit the computation time, so that the method gives what we call
a second approximation.

\begin{figure}[tbph]
\begin{centering}
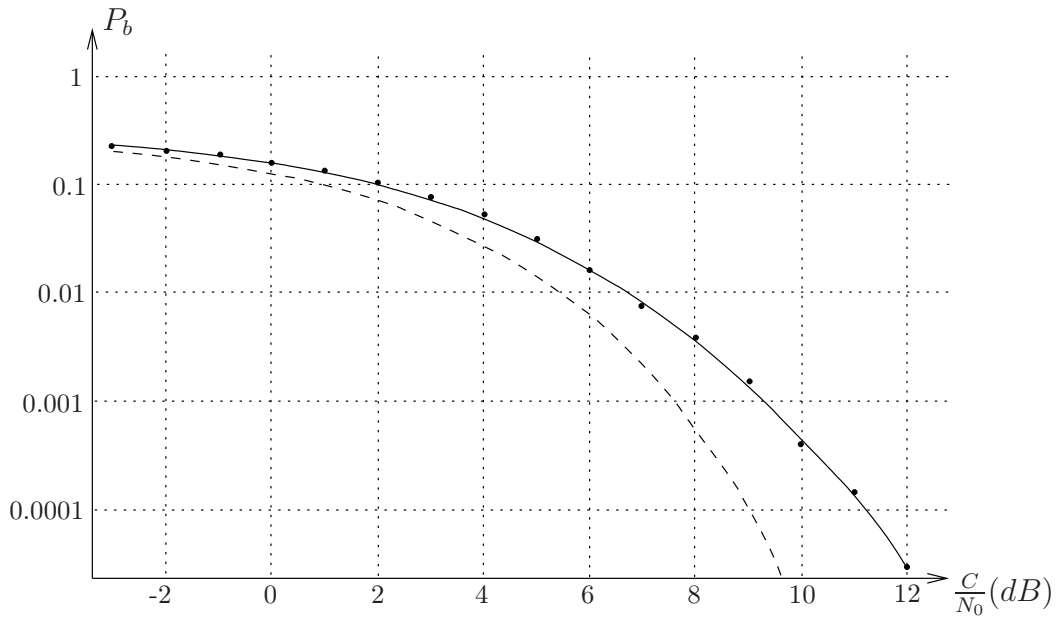
\par\end{centering}
\caption{Comparison of methods giving the binary error-rate for CQPRS with
a non-linear distortion in a traveling-wave tube, as a function of
$C/N_{0}$: theory without distortion (dashed line), second theoretical
approximation by recalculation of the survivors (full line), and simulations
(points). \label{figure22}}
\end{figure}

Our results are drawn in Figure~\ref{figure22} (the saturation level
$A_{s}$ was chosen equal to $3$). This figure shows a good agreement
between theory and simulation.

\subsection{COQPRS modulation with \viterbi decoding, no precoding and unmatched
filtering}

For this modulation, where one of the two duobinary streams is off-set
by $T_{s}/2$, we will only use the method described as the second
approximation by reconstruction of the survivors. The calculation
is similar to that of the previous section, except that $x_{i}(kT_{s})$
can now take a number of values corresponding to the possible level
of the duobinary eye diagram at half of a symbol period from the sampling
instant. These values depend on the whole duobinary sequence, which
is here the interfering sequence.

We have used the reconstructed eye diagram where all the levels were
computed (by summation of a large number of impulse responses) and
stored in a computer file, for all duobinary sequences of length $9$.

To obtain the value of $x_{i}(kT_{s})$ for a given interfering sequence,
one has to search the sequence of length $9$ identical on $9$ symbols
to the interfering sequence, the symbol of order $i$ of this sequence
corresponding to the middle symbol (of order $5$) of the sequence
of length $9$. The level of the eye diagram at $T_{s}/2$ is then
read in the file, at the appropriate address, and divided by $\sqrt{2}$
to obtain $x_{i}(kT_{s})$. The values of the duobinary symbols recognized
by the decoder $z_{r}(kT_{s})$, of the interferences $y(i,j)$, the
survivors with and without distortion and finally the binary errors
are then obtained as for CQPRS, but with a larger number of calculations
and a longer computation time.

\begin{figure}[tbph]
\begin{centering}
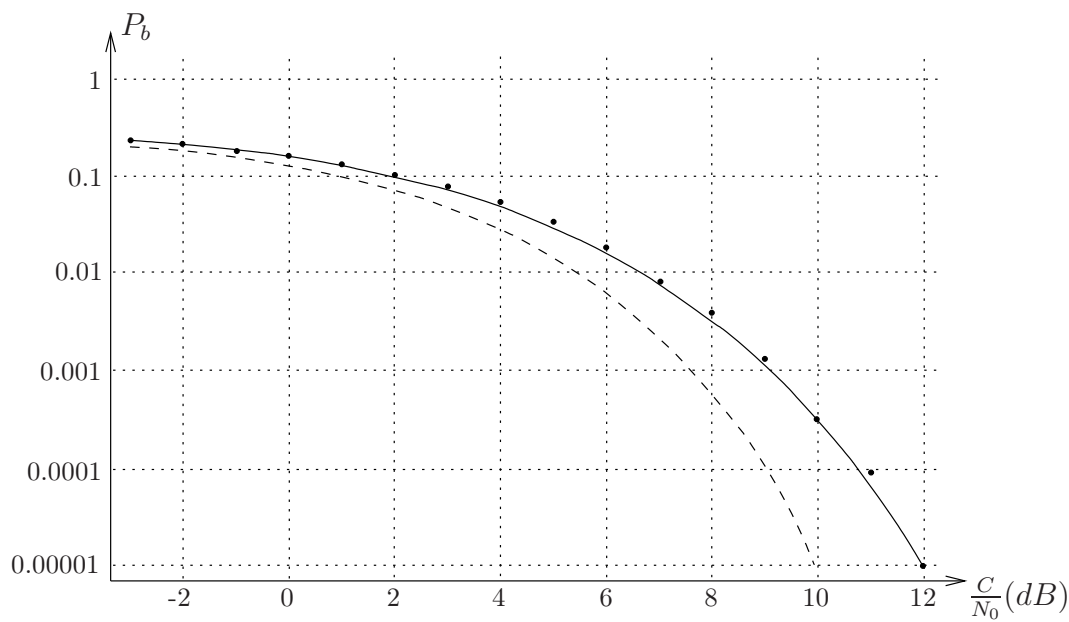
\par\end{centering}
\caption{Comparison of methods giving the binary error-rate for COQPRS with
a non-linear distortion in a traveling-wave tube, as a function of
$C/N_{0}$: theory without distortion (dashed line), second theoretical
approximation by recalculation of the survivors (full line), and simulations
(points).\label{figure23}}
\end{figure}

The results are given in Figure~\ref{figure23} where it is seen
that there is, again, good agreement between theory and simulation.

In terms of $C/N_{0}$, COQPRS seems to be slightly better than CQPRS
since the degradation due to the non-linear distortion is $1.7\,\decibel$
for the first, and $2.1\,\decibel$ for the second one at a BER of
$10^{-4}$. Moreover for these two modulations, \viterbi decoding
still provides an improvement of about $1.5$ to $2\,\decibel$ over
threshold decoding.

\section{Conclusions}

The duobinary code, which is the subject of the present study, is
used in a number of communications and broadcasting systems. The binary
error-rate (BER) is normally evaluated by simulation with pseudo-random
binary sequences as input data. In all cases, decoding of the duobinary
code with the \viterbi  algorithm provides a significant improvement
with respect to the simple threshold decoding.

With \viterbi decoding, the theoretical value of the BER is often
not known, specially for channels strongly impaired by noise and linear
or non-linear distortions. In the previous Sections, several numerical
methods usable on a simple computer were presented for the calculation
of the theoretical BER, in presence of noise and typical linear or
non-linear distortions. In all cases, the results obtained were in
good agreement with those of simulation.

\subsubsection*{Special note}

This paper is published in memory of Professor Henri Mertens. Henri
Mertens was Associate Professor Emeritus at the University of Liège
(Liège), Belgium, and also former Deputy Director of the Technical
Center of the European Broadcasting Union (EBU), Brussels, Belgium.

\settowidth{\nomlabelwidth}{CQPRS}
\printnomenclature{}


\end{document}